\documentclass{gCOV2e}

\theoremstyle{plain}
\newtheorem{theorem}{Theorem}[section]

\theoremstyle{remark}
\newtheorem{remark}[theorem]{Remark}

\begin{document}
\jvol{00} \jnum{00} \jyear{2013} \jmonth{August}

\title{Improved algorithm for analytical solution of the heat conduction \break problem  in doubly
periodic 2D composite materials}

\author{D. Kapanadze$^{a}$, G. Mishuris$^{b}$  and E. Pesetskaya$^{a}$$^{\ast}$\thanks{$^\ast$Corresponding author. Email: kate.pesetskaya@gmail.com
\vspace{6pt}}\\\vspace{6pt}  $^{a}${\em{A. Razmadze Mathematical Institute, Tbilisi State University, Georgia}};
$^{b}${\em{Institute of Mathematics and Physics, Aberystwyth
University, UK}}\\\vspace{6pt}\received{August 2013} }

\maketitle

\begin{abstract}
We consider a boundary value problem (BVP)  in unbounded 2D doubly periodic composite
with circular inclusions having arbitrary constant conductivities.
By introducing complex potentials, the BVP
for the Laplace equation is transformed to a special $R$-linear BVP for doubly periodic analytic functions.
This problem is solved with use of the method of functional equations. The $R$-linear BVP is
transformed to a system of functional equations.  A new improved algorithm for solution of the system is proposed.
It allows one not only to compute the average property but
to reconstruct the solution components (temperature and flux) at an arbitrary point of the composite.
Several computational examples are discussed in details demonstrating high efficiency
of the method. Indirect estimate of the algorithm accuracy has been also provided.
\end{abstract}


\begin{classcode}
30E25, 35B27, 74Q05, 33E05
\end{classcode}\medskip

\section{Introduction}

Heterogeneous media model
problems  serve the purposes of material science studies for
the analysis of the various fields and prediction of their properties \cite{BeMi01,Ki72,Markov,SuWoYo}.
Different approaches for study linear inhomogeneous material are presented in well-known
monographs \cite{Alla,Cher,KalKol,Manevitch,Mi02,MovMovPou}. One of the
approaches dealing with composite materials is the so-called
homogenization method (see \cite{BakPan,JKO}).
Mathematical aspects of the higher order homogenization have been
 developed in \cite{CheSmy04}. The limiting
case for large (close to the maximal value) rectangular
cross-section cylindrical cavities by means of an asymptotic
procedure were studied in \cite{BakPan} where explicit analytical
expressions for effective parameters have been also found.
Non-local phenomena resulting from a high contrast (or anisotropy)
of composite structures were studied in \cite{Alla,CheSmyZhi06}. In two- and three-dimensional cases  the
Rayleigh multipole expansions method and its generalizations is
effectively used (see, e.g.\cite{MovMovPou,Kusch}). Various analytical
approaches have been discussed in \cite{MiR,Obn}.

Essential progress has been already achieved in the area of numerical analysis of
composite material properties. Such approach is naturally
restricted to a finite cell (or a few cell - representative element) size in order
to reduce the computation cost. A vast literature related to this approach can be found in \cite{Zohdi}.
The major advantage of analytical approach is a possibility to describe and analyze the material properties
by means of explicit analytical formulas. This allow one to reveal an influence of the materials
characteristics (like size, shape, location of components,
their material properties) on the overall properties of the composite (homogeneous approach) \cite{Andr3,BeMi01,FPOG,Kach,Kanaun_Levin,MitPesRog,Pes05}. Recently, the relationship between the effective properties in the problem of the heat conduction and elasticity have been revealed and effectively exploited \cite{Kach_Sev,Sev_Kach}.

In this paper, we reveal another advantage of the analytical approach showing that it is capable to efficiently reconstruct the global and local distributions of the physical fields. We consider well-known linear heat conduction problem in 2D unbounded doubly periodic composite with material properties independent of the temperature field.
The components (inclusions) are supposed to be disjoint disks formed a doubly periodic structure.
We consider the steady process governed here by the Laplace equation. We will mostly follow by pioneering
work \cite{BeMi01}, but a few important improvements will be introduced.
First, we slightly change the problem formulation introducing more natural conditions at infinity prescribing only average flux, at an arbitrary direction, in contrast to the problem investigated in \cite{BeMi01},
where a special temperature distribution assumed in the direction of the coordinate system.
In the linear formulation, our approach is in fact equivalent to periodic conditions for flux on the boundary of the minimal representative cell.

Although, we treat the problem using the methods developed in \cite{BeMi01}, reducing the corresponding BVP to a system of functional equations with respect to doubly periodic analytical functions, we substantially change the algorithm for the numerical calculations. The algorithm in \cite{BeMi01} is mostly oriented to find the effective conductivity. In this case, it is sufficient to know values of the heat flux in the centers of inclusions only. However, it turns out that it may not guarantee the best accuracy
when defining values of the flux outside the inclusion or reconstructing the temperature distribution.
Modified algorithm presented in this work allows one to increase an accuracy of the numerical computations an any point
in the distance from the centers of inclusions and {to find the temperature field (with accuracy to an arbitrary constant).
The proposed modification allows us to find the flux distribution in an explicit form containing all parameters of the
considered model such as conductivities of the matrix and inclusions,
radii and centers of inclusions, an intensity and an angle of the flux. As the previous algorithm, it also relays on
the values of special Eisenstein functions (\cite{Weil}).

The paper is organized as follows. In Section 2 we describe the
geometry of the considered composites and formulate the mathematical problem
basing on proper physical assumptions. In Section 3 we briefly overview the auxiliary problem
stated in \cite{BeMi01}, show a connection with the original problem.

In Section 4 we describe a new algorithm in details and show that both components of the solution, the flux and the temperature, can be computed in the unique scheme. Finally, numerical calculations are performed and discussed in Section 5 to demonstrate algorithm accuracy, robustness and effectiveness.

\section{Statement of the problem}\label{sec:2}
We consider a lattice $L$ which is defined by the two
fundamental translation vectors ``$1$'' and ``$\imath$'' (where
$\imath^2=-1$) in the complex plane $\mathbb{C}\cong \mathbb{R}^2$
(with the standard notation $z=x+\imath y$).

\begin{figure}
\begin{center}
\resizebox*{10cm}{!}{\includegraphics{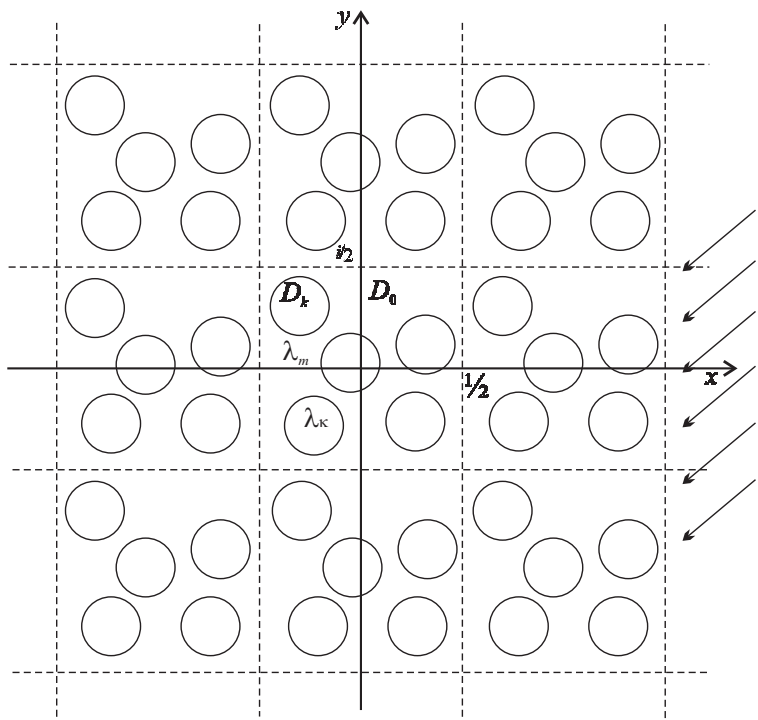}}%
\caption{The representative cell $Q_{(0,0)}$ within doubly periodic composite.}%
\label{fig:1}
\end{center}
\end{figure}

Here, the
representative cell (see Figure~\ref{fig:1}) will be the square
\begin{equation}
Q_{(0,0)} := \left\{ z= t_1+\imath t_2 \in \mathbb{C}:
-\frac{1}{2}<t_{p}<\frac{1}{2},\, p=1,2\right\}.
\end{equation}
 Let $\mathcal{E}:=
\bigcup\limits_{m_1, m_2} \{m_1+\imath m_2\}$ be the set of the
lattice points, where $m_1, m_2 \in \mathbb{Z}$. The cells
corresponding to the points of the lattice $\mathcal{E}$ will be
denoted by
\begin{equation}
Q_{(m_1,m_2)}=Q_{(0,0)}+m_1+\imath m_2:= \left\{z\in \mathbb{C}:
z-m_1-\imath m_2 \in Q_{(0,0)}\right\}. \end{equation}
 It is considered the situation
when mutually disjoint disks (inclusions)
$D_k:=\{z \in \mathbb{C}: |z-a_{k}|<r_k\}$ with different radii $r_k$ and boundaries
$\partial D_k:=\{z \in \mathbb{C}: |z-a_{k}|=r_k\} \, (k=1,2,\dots,N)$
are located inside the cell $Q_{(0,0)}$ and periodically repeated
in all cells $Q_{(m_1,m_2)}$. Let us denote by
\begin{equation}
D_0:=Q_{(0,0)}\setminus \left(\bigcup\limits_{k=1}^{N}\,D_k\cup
\partial D_k\right)
\end{equation}
the connected domain obtained by removing of the inclusions from
the cell $Q_{(0,0)}$ (cf.\ Figure~\ref{fig:1}).

Let us consider the problem of determination of the heat flux function of a doubly periodic composite material with matrix
\begin{equation}
D_{matrix}=\bigcup\limits_{m_1,m_2} \,((D_0 \cup \partial
Q_{(0,0)})+m_1+\imath m_2) \end{equation}
and inclusions
\begin{equation}
D_{inc}=\bigcup\limits_{m_1,m_2} \bigcup\limits_{k=1}^{N}\,
(D_k+m_1+\imath m_2) \end{equation} occupied by materials of
conductivities $\lambda_m > 0$ and $\lambda_k > 0$ ($k=1,\dots,N$), respectively.
For this purpose, we consider a problem of the determination of the potential
of the corresponding fields, i.e.,  a temperature function $T=T(x,y)$ satisfying the
Laplace equation in each component of the composite material
\begin{equation}\label{Lapl}
\Delta T(z)=0,\qquad z \in D_{matrix}\cup D_{inc},
\end{equation}
which have to satisfy the following boundary conditions on all
$\partial D_k, \,k=1,2,\dots,N$:
\begin{equation}\label{bound_1}
T(t)=T_k(t),
\end{equation}
\begin{equation}\label{bound_2}
\lambda_m \frac{\partial T}{\partial n}(t)=\lambda_k
\frac{\partial T_k}{\partial n}(t), \qquad t \in \bigcup\limits_{m_1,m_2}\,\partial D_k.
\end{equation}
Here, the vector $n=(n_1,n_2)$ is the outward unit
normal vector to $\partial D_k$,
$\frac{\partial}{\partial n}=n_1
\frac{\partial}{\partial x}+n_2 \frac{\partial}{\partial y}$
is the outward normal derivative, and
\begin{equation}
T(t):=\lim\limits_{z \rightarrow t, z \in D_0}\, T(z), \qquad
T_k(t):=\lim\limits_{z \rightarrow t, z \in D_k}\, T(z).
\end{equation}
The conditions (\ref{bound_1})--(\ref{bound_2}) form the so-called {\it
ideal (perfect) contact conditions}.

The thermal loading for the composite is described weakly by the flux
given at infinity or more accurately by its intensity $A$. We assume that the flux
is directed $\theta$ which does not coincide, in general, with the
orientation of the periodic cell (see, Figure~\ref{fig:1}). According to the
conservation law and the ideal contact condition between
the different materials, the flux is continuous in the entire
structure. Moreover, as a result of such formulation, the
temperature, which is also continuous as the results of the ideal transmission conditions along the interface between the matrix and inclusions,
possesses non-zero jumps across any cell.

In addition, we assume that the heat flux
is periodic on $y$. Thus,
\begin{equation}\label{bound_cell_1}
\lambda_m T_y\Bigl(x,\frac{1}{2}\Bigr)=\lambda_m T_y\Bigl(x,-\frac{1}{2}\Bigr)=-A
\sin \theta+q_1(x),
\end{equation}
where $A$ is the intensity of an external flux. The heat flux is
periodic on $x$, consequently,
\begin{equation} \label{bound_cell_2}
\lambda_m T_x\Bigl(-\frac{1}{2},y\Bigr)=\lambda_m T_x\Bigl(\frac{1}{2},y\Bigr)=-A
\cos \theta+q_2(y).
\end{equation}
To complement to the average flux conditions at infinity, the latter
immediately proves that the equalities
\begin{equation} \label{zero_flux}
\int\limits_{-1/2}^{1/2}q_j(\xi)d\xi=0
\end{equation}
are valid for the unknown functions $q_j$, ($j=1,2$). As a result
of (\ref{bound_cell_1}) and (\ref{bound_cell_2}), the heat flux has a zero
mean value along the cell
\begin{equation}\label{bound_4aa}
\int\limits_{\partial\,
Q_{(m_1,m_2)}}\hspace{-5mm}\frac{\partial
T(s)}{\partial n}ds=0.
\end{equation}
From the physics point of view, condition (\ref{bound_4aa}) is the consequence of the fact that no source (sink) exists in the cells.
Moreover, since there is no source (sink)
inside the composite, i.e., neither in the matrix of the composite,
nor in any inclusion (the total heat flux through any closed
simply connected curve is equal to zero), we have
\begin{equation}\label{bound_4aa0}
\int\limits_{\partial D_k +m_1+\imath
 m_2}\hspace{-5mm}\frac{\partial T}{\partial n}ds=0.
\end{equation}

We will introduce complex potentials $\varphi(z)$ and
$\varphi_k(z)$ which are analytic in $D_0$ and $D_k$, and
continuously differentiable in the closures of $D_0$ and $D_k$,
respectively, by using the following relations
\begin{equation}\label{potential_1}
T(z)= \left\{
\begin{array}{lc} {\mathrm{Re}}\,(\varphi(z)+B z), \; z \in D_{matrix},
\\  \\  \frac{2\lambda_m}{\lambda_m+\lambda_k}\,{\mathrm{Re}}\, \varphi_k(z), \; z \in
D_{inc},
\end{array}
\right.
\end{equation}
where $B$ is an unknown constant belong to $\mathbb{C}$. Besides, we assume the the real part of $\varphi$ is doubly periodic in $D_0$, i.e.
$$
\mathrm{Re}\,\varphi(z+1)-\mathrm{Re}\,\varphi(z)= 0, \qquad \mathrm{Re}\,\varphi(z+\imath)-\mathrm{Re}\,\varphi(z)=0.
$$
Note that in general the imaginary part of $\varphi$ is not doubly periodic in $D_0$.

 Let us show that $\varphi$ is single-valued function in $D_{matrix}$.
We take a harmonic function $v$ in  $D_{matrix}$
which is the harmonic conjugate to $T$. For this pair of functions the
Cauchy-Riemann equations $\frac{\partial T}{\partial
x}=\frac{\partial v}{\partial y}$, $\frac{\partial T}{\partial
y}=-\frac{\partial v}{\partial x}$ (or the so called
normal-tangent Cauchy-Riemann equations $\frac{\partial
T}{\partial n}=\frac{\partial v}{\partial s}$, $\frac{\partial
T}{\partial s}=- \frac{\partial v}{\partial n}$) have to be valid.
The functions $v$ has the following form:
\begin{equation}\label{potential_2}
v(z)= \left\{
\begin{array}{lc} {\mathrm{Im}}\,(\varphi(z)+B z), \; z \in D_{matrix},
\\  \\  \frac{2\lambda_m}{\lambda_m+\lambda_k}\,{\mathrm{Im}}\, \varphi_k(z), \; z \in
D_{inc},
\end{array}
\right.
\end{equation}
with the same unknown constant $B$.

As it follows from (\ref{bound_4aa}) -- (\ref{bound_4aa0}) we can write
\begin{equation}\label{bound_4a_v1}
\int\limits_{\partial\, Q_{(m_1,m_2)}}\hspace{-5mm}\frac{\partial
v}{\partial s}ds=0,\quad \int\limits_{\partial D_k +m_1+\imath
 m_2}\hspace{-5mm}\frac{\partial v}{\partial s}ds=0.
\end{equation}
These relations yield that the harmonic function $v$ is
single-valued in the domain $D_{matrix}$. Therefore, the complex potential $\varphi(z)$ is single-valued
 in $D_{matrix}$.

To determine the flux $\nabla T(x,y)$, we need to obtain
derivatives of the complex potentials:

\begin{equation}\label{flux_funs}
\begin{array}{ll}
\psi(z):=\frac{\partial \varphi}{\partial z}=\frac{\partial
T}{\partial x}-\imath\frac{\partial T}{\partial y}-B,\,
& z \in D_0,\\
\ &\ \\
\psi_k(z):=\frac{\partial \varphi_k}{\partial
z}=\frac{\lambda_m+\lambda_k}{2\lambda_m}\left(\frac{\partial
T_k}{\partial x}-\imath\frac{\partial T_k}{\partial y}\right),
& z \in D_k.
\end{array}
\end{equation}

Let us rewrite conditions (\ref{bound_1})--(\ref{bound_2}) in terms of the
complex potentials $\varphi(z)$ and $\varphi_k(z)$.
Let $s$ be the natural parameter of the curve $\partial D_k$ and
\begin{equation}
\frac{\partial}{\partial s}=-n_2\frac{\partial}{\partial
x}+n_1\frac{\partial}{\partial y}
\end{equation}
be the tangent
derivative along $\partial D_k$. Applying the Cauchy-Riemann equations, the equality (\ref{bound_2}) can be written as
\begin{equation}
\lambda_m \frac{\partial v}{\partial s}(t)=\lambda_k
\frac{\partial v_k}{\partial s}(t), \qquad |t-a_k|=r_k.
\end{equation}
Integrating the last equality on $s$, we arrive at the
relation
\begin{equation}
\lambda_m v(t)=\lambda_k v_k(t)+c,
\end{equation}
where $c$ is an arbitrary constant. We put $c=0$ since the
imaginary part of the function $\varphi$ is determined up to an
additive constant which does not impact on the form of $T$. Using
(\ref{potential_2}), we have
\begin{equation}\label{8}
{\mathrm{Im}} \,\varphi(t)=-{\mathrm{Im}} \,(B t)+\frac{2\lambda_k}{\lambda_m+\lambda_k} {\mathrm{Im}}\,
\varphi_k(t), \qquad |t-a_k|=r_k.
\end{equation}

Using (\ref{potential_1}), we are able to write the equality (\ref{bound_1}) in the
following form:
\begin{equation}\label{9}
{\mathrm{Re}}\, \varphi(t)=-{\mathrm{Re}}\,(B t)+\frac{2
\lambda_m}{\lambda_m+\lambda_k} {\mathrm{Re}}\, \varphi_k.
\end{equation}
Adding the relation (\ref{9}) and (\ref{8}) multiplied by $\imath$,
and using ${\mathrm{Re}}\, \varphi_k=\frac{\varphi_k+\overline{\varphi_k}}{2},
\, {\mathrm{Im}}\, \varphi_k=\frac{\varphi_k-\overline{\varphi_k}}{2 \imath}$,
$t-a_k=\frac{r_k^2}{\overline{t-a_k}}$, we have
\begin{equation} \label{10}
\varphi(t)=\varphi_k(t)-\rho_k \overline{\varphi_k(t)}-B t, \qquad |t-a_k|=r_k,
\end{equation}
where $\rho_k=\frac{\lambda_k-\lambda_m}{\lambda_k+\lambda_m}$.

Let us now differentiate (\ref{10}). First, note that
\begin{equation}
[\overline{\varphi(t)}]'=-\left(\frac{r_k}{t-a_k}\right)^2\,
\overline{\varphi'(t)},\qquad |t-a_k|=r_k.
\end{equation}
This can be
easily shown by representing the function $\varphi$ in the form
$\varphi(z)=\sum\limits_{l=0}^{\infty}\,\alpha_k (z-a_k)^l$,
$|z-a_k|\leq r_k$, and by using the relation
$t=\frac{r_k^2}{\overline{t-a_k}}+a_k$ on the boundary
$|t-a_k|=r_k$. Thus, after differentiating (\ref{10}) and using (\ref{flux_funs}), we arrive
at the following {\it $\mathbb{R}$-linear conjugation
problem} (\cite{MiR}) on each contour $|t-a_k|=r_k$,
\begin{equation} \label{11}
\psi(t)=\psi_k(t)+\rho_k
\left(\frac{r_k}{t-a_k}\right)^2\overline{\psi_k(t)}-B,
\end{equation}
with  $k=1,2,\dots,N$.
\begin{remark}
Thus, the boundary value problem (\ref{Lapl})-(\ref{bound_2}), (\ref{bound_cell_1})-(\ref{bound_cell_2}) for harmonic functions is reduced to  $\mathbb{R}$-linear conjugation problem (\ref{11}) for analytical doubly periodic functions $\psi, \psi_1, \dots,\psi_N$.
\end{remark}

We will seek a solution $\psi(z), \psi_k(z)$ of the problem (\ref{11}) as a sum  $\psi(z)=\psi^{(1)}(z)+\psi^{(2)}(z)$,
$\psi_k(z)=\psi_k^{(1)}(z)+\psi_k^{(2)}(z)$ of solutions of the following two BVPs:
\begin{equation}\label{psi_1}
\psi^{(1)}(t)=\psi^{(1)}_k(t)+\rho_k
\left(\frac{r_k}{t-a_k}\right)^2\overline{\psi^{(1)}_k(t)}-B_1,
\end{equation}
\begin{equation}\label{psi_2}
\psi^{(2)}(t)=\psi^{(2)}_k(t)+\rho_k
\left(\frac{r_k}{t-a_k}\right)^2\overline{\psi^{(2)}_k(t)}-\imath B_2,
\end{equation}
where $\psi_k^{(1)}$ and $\psi_k^{(2)}$ are analytical doubly periodic functions, $B=B_1+\imath B_2$.

\section{Formulation of an auxiliary problem}\label{sec:3}

 In this section we briefly overview the auxiliary problem discussed in \cite{BeMi01} and represent necessary results in convenient for us form.
Let $\widetilde{T}$ be a solution of the boundary value problem (\ref{Lapl})-(\ref{bound_2}) with a constant jump corresponding to the external field applied in the $x$-direction
\begin{equation}\label{jump}
\widetilde{T}(z+1)=\widetilde{T}(z)+1,\quad \widetilde{T}(z+\imath)=\widetilde{T}(z).
\end{equation}

The complex potentials $\widetilde{\varphi}^{(1)}(z)$ and
$\widetilde{\varphi}^{(1)}_k(z)$ are introduced as follows
\begin{equation}\label{potential_1t}
\widetilde{T}(z)= \left\{
\begin{array}{lc} {\mathrm{Re}}\,(\widetilde{\varphi}^{(1)}(z)+z), \; z \in D_{matrix},
\\  \\  \frac{2\lambda_m}{\lambda_m+\lambda_k}\,{\mathrm{Re}}\, \widetilde{\varphi}^{(1)}_k(z), \; z \in
D_{inc}.
\end{array}
\right.
\end{equation}
Note that $\widetilde{\varphi}^{(1)}(z)$ and
$\widetilde{\varphi}^{(1)}_k(z)$  are analytic in $D_0$ and $D_k$, and
continuously differentiable in the closures of $D_0$ and $D_k$,
respectively. Besides, the real part of $\widetilde{\varphi}^{(1)}$ is doubly periodic in $D_0$, i.e.
\begin{equation}\label{doubl_per}
{\mathrm{Re}}\,\widetilde{\varphi}^{(1)}(z+1)-{\mathrm{Re}}\,\widetilde{\varphi}^{(1)}(z)= 0, \qquad {\mathrm{Re}}\,\widetilde{\varphi}^{(1)}(z+\imath)-{\mathrm{Re}}\,\widetilde{\varphi}^{(1)}(z)=0.
\end{equation}
In general, the imaginary part of $\widetilde{\varphi}^{(1)}$ is not doubly periodic in $D_0$. It turns out that they satisfy the following $\mathbb{R}$-linear conjugation boundary value problem obtained in \cite{BeMi01}:
\begin{equation} \label{10_Mit}
\widetilde{\varphi}^{(1)}(t)=\widetilde{\varphi}^{(1)}_k(t)-\rho_k \overline{\widetilde{\varphi}^{(1)}_k(t)}-t, \qquad |t-a_k|=r_k.
\end{equation}
Differentiating  the last equality, we obtain that the boundary value problem (\ref{Lapl})-(\ref{bound_2}), (\ref{jump}) is reduced to the $\mathbb{R}$-linear conjugation boundary value problem for analytical doubly periodic functions
$\widetilde{\psi}^{(1)}, \widetilde{\psi}^{(1)}_1, \dots,\widetilde{\psi}^{(1)}_N$ (cf.\ \cite{BeMi01}):
\begin{equation}\label{11_Mit}
\widetilde{\psi}^{(1)}(t)=\widetilde{\psi}^{(1)}_k(t)+\rho_k
\left(\frac{r_k}{t-a_k}\right)^2\overline{\widetilde{\psi}^{(1)}_k(t)}-1
\end{equation}
with
\begin{equation}\label{potential_Mit}
\frac{\partial \widetilde{T}}{\partial x}-\imath \frac{\partial \widetilde{T}}{\partial y}= \left\{
\begin{array}{lc} \widetilde{\psi}^{(1)}(z)+ 1, \; z \in D_{matrix},
\\  \\  \frac{2\lambda_m}{\lambda_m+\lambda_k}\, \widetilde{\psi}^{(1)}_k(z), \; z \in
D_{inc},
\end{array}
\right.
\end{equation}
and
\begin{equation}\label{flux_funs_aux}
\begin{array}{ll}
\widetilde{\psi}^{(1)}(z):=\frac{\partial \widetilde{\varphi}^{(1)}}{\partial z}=\frac{\partial
\widetilde{T}}{\partial x}-\imath\frac{\partial \widetilde{T}}{\partial y}-1,\,
& z \in D_0,\\
\ &\ \\
\widetilde{\psi}^{(1)}_k(z):=\frac{\partial \widetilde{\varphi}^{(1)}_k}{\partial
z}=\frac{\lambda_m+\lambda_k}{2\lambda_m}\left(\frac{\partial
\widetilde{T}_k}{\partial x}-\imath\frac{\partial \widetilde{T}_k}{\partial y}\right),
& z \in D_k.
\end{array}
\end{equation}

Besides, we mention that when the temperature has a constant jump corresponding to the external field applied in the $y$-direction
\begin{equation*}
\widetilde{T}(z+1)=\widetilde{T}(z),\quad \widetilde{T}(z+\imath)=\widetilde{T}(z)-1,
\end{equation*}
the temperature is defined as
\begin{equation}\label{potential_1t_90}
\widetilde{T}(z)= \left\{
\begin{array}{lc} {\mathrm{Re}}\,(\widetilde{\varphi}^{(2)}(z)+\imath z), \; z \in D_{matrix},
\\  \\  \frac{2\lambda_m}{\lambda_m+\lambda_k}\,{\mathrm{Re}}\, \widetilde{\varphi}^{(2)}_k(z), \; z \in
D_{inc},
\end{array}
\right.
\end{equation}
with corresponding functions $\widetilde{\varphi}^{(2)}, \widetilde{\varphi}_k^{(2)}$ possess the same properties as the functions $\widetilde{\varphi}^{(1)}, \widetilde{\varphi}_k^{(1)}$, and
$$
\widetilde{\varphi}^{(2)}(t)=\widetilde{\varphi}^{(2)}_k(t)-\rho_k \overline{\widetilde{\varphi}^{(2)}_k(t)}-\imath t, \qquad |t-a_k|=r_k.
$$
The corresponding $\mathbb{R}$-linear conjugation boundary value problem has a form
\begin{equation}\label{11_Mit2}
\widetilde{\psi}^{(2)}(t)=\widetilde{\psi}^{(2)}_k(t)+\rho_k
\left(\frac{r_k}{t-a_k}\right)^2\overline{\widetilde{\psi}^{(2)}_k(t)}-\imath, \quad |t-a_k|=r_k.
\end{equation}

The problems (\ref{psi_1}) and (\ref{psi_2}) can be reduced to the problems (\ref{11_Mit})  and (\ref{11_Mit2}) by the
following replacements:
\begin{equation}\label{repl1}
\psi^{(1)}(z)=B_1 \widetilde{\psi}^{(1)}(z), \quad \psi^{(1)}_k(z)=B_1 \widetilde{\psi}^{(1)}_k(z),
\end{equation}
\begin{equation}\label{repl2}
\psi^{(2)}(z)=B_2 \widetilde{\psi}^{(2)}(z), \quad \psi^{(2)}_k(z)=B_2 \widetilde{\psi}^{(2)}_k(z).
\end{equation}

\begin{remark}\label{rem:i}
It is easy to verify that the functions $\widetilde{\psi}^\bot(z):=\imath \widetilde{\psi}^{(2)}(\imath z)$ and $\widetilde{\psi}^\bot_k(z):=\imath \widetilde{\psi}^{(2)}_k(\imath z)$ satisfy the following $\mathbb{R}$-linear conjugation boundary value problem
\begin{equation}\label{11_Mitbot}
\widetilde{\psi}^\bot(t)=\widetilde{\psi}^\bot_k(t)+\rho_k
\left(\frac{r_k}{t-b_k}\right)^2\overline{\widetilde{\psi}^\bot_k(t)}+1, \quad |t-b_k|=r_k,\ \  b_k=-\imath a_k.
\end{equation}
\end{remark}
Note that $\widetilde{\psi}^{(2)}(z)=-\imath\widetilde{\psi}^\bot(-\imath z)$ and
$\widetilde{\psi}^{(2)}_k(z)=-\imath\widetilde{\psi}^\bot_k(-\imath z)$.

Thus, to find a solution of the problem (\ref{11}), it is sufficient to find solutions $\widetilde{\psi}^{(1)}(z), \widetilde{\psi}_k^{(1)}(z)$ and $\widetilde{\psi}^\bot(z), \widetilde{\psi}_k^\bot(z)$ of the problems (\ref{11_Mit}) and (\ref{11_Mitbot}), respectively.

%%%%%%%%%%%%%%%%%%%%%%%%%%%%%%%%%%%%%%%%%%%%%%%%%%%%%%%%%%%%%%%%%%%%%%%%%%%%%%%%%%%%%%%%%%%%%%%%%%%%%%%%%%%%%%%%%%%%%%%%%%%%%%%%%%%%%%%%%%%%%%%%%%%%%%%%%%%
\section{Solution of the problem}\label{sec:4}

First, let us find the real constants $B_1$ and $B_2$.

We introduce further notations
\begin{eqnarray*}
I:=\int\limits_{-\frac{1}{2}}^{\frac{1}{2}}\,{\mathrm{Re}}\,\widetilde{\psi}^{(1)}\Bigl(\frac{1}{2}+\imath y\Bigl)\,dy, \quad
I^\bot:=\int\limits_{-\frac{1}{2}}^{\frac{1}{2}}\,{\mathrm{Re}}\,\widetilde{\psi}^\bot\Bigl(\frac{1}{2}+\imath y\Bigl)\,dy.
\end{eqnarray*}
In general, the integrals $I$ and $I^\bot$ differ from zero. As it is shown in  Remark \ref{rem:I21} below
\begin{equation}\label{I21}
\int\limits_{-\frac{1}{2}}^{\frac{1}{2}}\,{\mathrm{Im}}\,\widetilde{\psi}^{(1)}\Bigl(x+\frac{\imath}{2}\Bigl)\,dx=0.
\end{equation}

Taking  $B_2=0$ (which corresponds to the problem (\ref{psi_1})) and using (\ref{potential_1}), (\ref{flux_funs}) and (\ref{repl1}), we obtain
$$
\lambda_m \frac{\partial T(x,\frac{1}{2})}{\partial y}=-\lambda_m {\mathrm{Im}}\,\Bigl(\psi\Bigl(x+\frac{\imath}{2}\Bigl)+ B\Bigl)=
-\lambda_m B_1 {\mathrm{Im}}\,\widetilde{\psi}^{(1)}\Bigl(x+\frac{\imath}{2}\Bigl),
$$
and
$$
\lambda_m \frac{\partial T(\frac{1}{2},y)}{\partial x}=\lambda_m {\mathrm{Re}}\,\Bigl(\psi\Bigl(\frac{1}{2}+\imath y\Bigl)+ B\Bigl)=
\lambda_m B_1\Bigl({\mathrm{Re}}\,(\widetilde{\psi}^{(1)}\Bigl(\frac{1}{2}+\imath y\Bigl)+ 1)\Bigl).
$$
Due to (\ref{I21}), integration on $\left[-\frac{1}{2},\frac{1}{2}\right]$ the first equality gives
$$
\lambda_m\int\limits_{-\frac{1}{2}}^{\frac{1}{2}}\, \frac{\partial T(x,\frac{1}{2})}{\partial y}dx=0.
$$
Integrating on $\left[-\frac{1}{2},\frac{1}{2}\right]$ the second equality and applying (\ref{bound_cell_2}), we obtain the constant $B_1$:
\begin{equation}\label{B1}
B_1=\frac{-A \cos\theta}{\lambda_m(I+1)}.
\end{equation}

Similarly, taking $B_1=0$ (which corresponds to the problem (\ref{psi_2})) and using (\ref{potential_1}), (\ref{flux_funs}) and (\ref{repl1}), we obtain
$$
\lambda_m \frac{\partial T(x,\frac{1}{2})}{\partial y}=-\lambda_m {\mathrm{Im}}\,\Bigl(\psi\Bigl(x+\frac{\imath}{2}\Bigl)+ B\Bigl)=
-\lambda_m B_2 {\mathrm{Im}}\,\Bigl(\widetilde{\psi}^{(2)}\Bigl(x+\frac{\imath}{2}\Bigl)\Bigl)-\lambda_m B_2,
$$
and
$$
\lambda_m \frac{\partial T(\frac{1}{2},y)}{\partial x}=\lambda_m {\mathrm{Re}}\,\Bigl(\psi\Bigl(\frac{1}{2}+\imath y\Bigl)+ B\Bigl)=
\lambda_m B_2{\mathrm{Re}}\,\Bigl(\widetilde{\psi}^{(2)}\Bigl(\frac{1}{2}+\imath y\Bigl)\Bigl).
$$
Using the equality $\widetilde{\psi}^{(2)}(z)=-\imath\widetilde{\psi}^\bot(-\imath z)$, we have
$$
{\mathrm{Im}}\,\Bigl(\widetilde{\psi}^{(2)} \Bigl(x+\frac{\imath}{2}\Bigl)\Bigl)=-{\mathrm{Re}}\, \Bigl(\widetilde{\psi}^\bot \Bigl(\frac{1}{2}-\imath x\Bigl)\Bigl),
$$
$$
{\mathrm{Re}}\,\Bigl(\widetilde{\psi}^{(2)} \Bigl(\frac{1}{2}+\imath y\Bigl)\Bigl)={\mathrm{Im}}\, \Bigl(\widetilde{\psi}^\bot \Bigl(-\frac{\imath}{2}+y\Bigl)\Bigl).
$$
Thus, we get
$$
\lambda_m\int\limits_{-\frac{1}{2}}^{\frac{1}{2}}\, \frac{\partial T(\frac{1}{2},y)}{\partial x}dy=0.
$$
Integrating on $\left[-\frac{1}{2},\frac{1}{2}\right]$  the term $\lambda_m \frac{\partial T(x,\frac{1}{2})}{\partial y}$ and applying (\ref{bound_cell_1}), we obtain the constant $B_2$:
\begin{equation}\label{B2}
B_2=\frac{-A \sin\theta}{\lambda_m(I^\bot-1)}.
\end{equation}

Taking into the account the properties of functions under consideration from Sections \ref{sec:2} and \ref{sec:3}, the results obtained above and  Remark \ref{rem:i}, we arrive at the following theorem:
\begin{theorem}\label{Temp_stat}
Let $T=T(x,y)$ and $T_k=T_k(x,y)$ be the solution of the problem (\ref{Lapl})-(\ref{bound_1}), (\ref{bound_cell_1}) and (\ref{bound_cell_2}). The temperature flux is defined in the following form:
\begin{equation} \label{Temp}
\frac{\partial T(x,y)} {\partial x}-\imath \frac{\partial T(x,y)} {\partial y}= \left\{
\begin{array}{lc} \psi(z)+B, \; z=x+\imath y \in D_{matrix},
\\  \\  \frac{2\lambda_m}{\lambda_m+\lambda_k}\,
\psi_k(z), \; z=x+\imath y \in D_{inc},
\end{array}
\right.
\end{equation}
where
$$
B=\frac{-A \cos\theta}{\lambda_m(I+1)}- \frac{A \sin\theta}{\lambda_m(I^\bot-1)}\, \imath,
$$
and
$$
\psi(z):=\frac{-A \cos\theta}{\lambda_m(I+1)}\widetilde{\psi}^{(1)}(z)+\imath \frac{A \sin\theta}{\lambda_m(I^\bot-1)}\widetilde{\psi}^\bot(-\imath z), \; z \in D_{matrix},
$$
$$
\psi_k(z):=\frac{-A \cos\theta}{\lambda_m(I+1)}\widetilde{\psi}^{(1)}_k(z)+\imath \frac{A \sin\theta}{\lambda_m(I^\bot-1)}\widetilde{\psi}_k^\bot(-\imath z), \; z \in D_{inc}.
$$
\end{theorem}

To find the temperature, it is sufficient to find the functions
$\varphi,\varphi_1,\dots,\varphi_N$ (cf. (\ref{potential_1})).
These functions can be represented as  sums $\varphi(z)=\varphi^{(1)}(z)+\varphi^{(2)}(z)$,
$\varphi_k(z)=\varphi_k^{(1)}(z)+\varphi_k^{(2)}(z)$ of two functions
$\varphi^{(1)}$ and $\varphi^{(2)}$ have to satisfy the following  BVPs:
\begin{equation}\label{phi_1}
\varphi^{(1)}(t)=\varphi^{(1)}_k(t)-\rho_k \overline{\varphi^{(1)}_k(t)}-B_1 t,
\end{equation}
\begin{equation}\label{phi_2}
\varphi^{(2)}(t)=\varphi^{(2)}_k(t)-\rho_k \overline{\varphi^{(2)}_k(t)}-\imath B_2 t.
\end{equation}
Analogously to (\ref{repl1})--(\ref{repl2}), we have $\varphi^{(1)}(z)=B_1 \widetilde{\varphi}^{(1)}(z)$, $\varphi^{(1)}_k(z)=B_1 \widetilde{\varphi}^{(1)}_k(z)$ and $\varphi^{(2)}(z)=B_2 \widetilde{\varphi}^{(2)}(z)$, $\varphi^{(2)}_k(z)=B_2 \widetilde{\varphi}^{(2)}_k(z)$.
It is easy to verify that
$$
\widetilde{\varphi}^{(2)}(z)=\widetilde{\varphi}^{(1)}(-\imath z), \; \widetilde{\varphi}^{(2)}_k(z)=\widetilde{\varphi}^{(1)}_k(-\imath z).
$$
The functions $\widetilde{\varphi}^{(1)}$ and $\widetilde{\varphi}^{(1)}_k$ can be found  up to an arbitrary constant as indefinite integrals of the functions $\widetilde{\psi}^{(1)}$ and $\widetilde{\psi}^{(1)}_k$, respectively (cf. (\ref{flux_funs_aux})).
Thus, we arrive at the following statement:
\begin{theorem} Let $T=T(x,y)$ and $T_k=T_k(x,y)$ be the solution of the problem (\ref{Lapl})-(\ref{bound_1}), (\ref{bound_cell_1}) and (\ref{bound_cell_2}). The temperature distribution can be found up to an arbitrary constant and is defined in the form (\ref{potential_1}), where
$$
B=\frac{-A \cos\theta}{\lambda_m(I+1)}- \frac{ A \sin\theta}{\lambda_m(I^\bot-1)}\, \imath,
$$
$$
\varphi(z)=\frac{-A \cos\theta}{\lambda_m(I+1)} \widetilde{\varphi}^{(1)}(z)-\frac{ A \sin\theta}{\lambda_m(I^\bot-1)} \widetilde{\varphi}^{(1)}(-\imath z),
$$
$$
\varphi_k(z)=\frac{-A \cos\theta}{\lambda_m(I+1)} \widetilde{\varphi}^{(1)}_k(z)-\frac{ A \sin\theta}{\lambda_m(I^\bot-1)} \widetilde{\varphi}^{(1)}_k(-\imath z).
$$
\end{theorem}

Now we describe a new algorithm for solution of the problem (\ref{11_Mit}). The problem (\ref{11_Mitbot}) can be solved analogously.
For convenience, we omit upper index in $\widetilde{\psi}^{(1)}$ and will write $\widetilde{\psi}$ below.
We shortly describe solvability of the problem (\ref{11_Mit}) using some facts and notation of the paper \cite{BeMi01}.

Notice that we have $N$ contours $\partial D_k$ and $N$ complex conjugation
conditions on each contour $\partial D_k$ but we need to find $N+1$ functions
$\widetilde{\psi}, \widetilde{\psi}_1, \dots,\widetilde{\psi}_N$. This means that we need one additional
condition to close up the system. For this  reason we introduce
a new doubly periodic function $\Phi$ which is a sectionally analytic in $Q_{(0,0)}$ and in
$\bigcup\limits_{k=1}^{N}\, D_k$ and has the zero jumps along each
$\partial D_k, \, k=1,2,\dots,N$.
Such consideration will give an additional condition on $\widetilde{\psi}, \widetilde{\psi}_1, \dots,\widetilde{\psi}_N$.
We will show that $\Phi \equiv 0$.

 Let us introduce the sectionally analytic doubly periodic function $\Phi$ by the
following formula:
\begin{equation}\label{sisPhi}
\Phi(z)=\left\{
\begin{array}{lc}
\widetilde{\psi}_k(z)- \sum\limits_{m=1}^{N} \rho_m
\sum\limits_{m_1,m_2}\hspace{-3mm}{}^{*}\,
W_{m_1,m_2,m}
\widetilde{\psi}_m(z)-1, \;
 |z-a_k|\leq r_k, \\
\widetilde{\psi}(z)-\sum\limits_{m=1}^{N}\rho_m\sum\limits_{m_1,m_2}\, W_{m_1,m_2,m}
\widetilde{\psi}_m(z), \; z \in D_0,
\end{array}
\right.
\end{equation}
 where
\begin{equation}\label{18}
W_{m_1,m_2,m} \widetilde{\psi}_m(z)=\left(\frac{r_m}{z-a_m-m_1-\imath
m_2}\right)^2 \overline{\widetilde{\psi}_m
\left(\frac{r_m^2}{\overline{z-a_m-m_1-\imath m_2}}+a_m \right)}
\end{equation}
and
\begin{equation}
\sum\limits_{m=1}^{N}\sum\limits_{m_1,m_2}\hspace{-2mm}{}^{*}\,
W_{m_1,m_2,m}:=\sum\limits_{m\neq k}\sum\limits_{m_1,m_2}\,
W_{m_1,m_2,m}+\sum\limits_{m_1,m_2}\hspace{-2mm}{}^{'}\,
W_{m_1,m_2,k}.
\end{equation}
The ``prime'' notation in $\sum\limits_{m_1,m_2}\hspace{-2mm}{}^{'}\,$
means
that the summation occurs in all $m_1$ and $m_2$ except at
$(m_1,m_2)=(0,0)$.

Applying \textit{Analytic Continuation Principle} and \textit{Liouville's} theorem for doubly periodic functions, we have that $\Phi=c$.

Let $\widetilde{\psi}$ and $\widetilde{\psi}_k$ be solutions of the
system $\Phi(z)=c$. Then, in $D_0$, we have
\begin{equation}
\widetilde{\psi}(z)=\widetilde{\psi}'(z)+c
\end{equation}
with some doubly periodic
function $\widetilde{\psi}'$. Inserting the last equality in (\ref{flux_funs_aux})
and then in (\ref{potential_1t}), we obtain
\begin{equation}
T(z)= {\mathrm{Re}}\,(\widetilde{\varphi}'(z)+cz+z), \quad z \in D_0,
\end{equation}
with some function $\widetilde{\varphi}'$ which
yields $c=0$. Thus, we have $\Phi(z)\equiv 0$.
Writing $\Phi(z) \equiv 0$, we obtain the
following  system of linear functional equations
\begin{equation}\label{sis1}
\widetilde{\psi}_k(z)= \sum\limits_{m=1}^{N}\rho_m \sum\limits_{m_1,m_2}\hspace{-2mm}{}^{*}\,
W_{m_1,m_2,m}
\widetilde{\psi}_m(z)+1
\end{equation}
which is uniquely solvable with respect to $\widetilde{\psi}_k$ in the space of analytical functions (for more details
cf. \cite{BeMi01}).

The function $\widetilde{\psi}$ has the form
\begin{equation}\label{eq1}
\widetilde{\psi}(z)=\sum\limits_{m=1}^{N}\rho_m \,\sum\limits_{m_1,m_2}\, W_{m_1,m_2,m}
\widetilde{\psi}_m(z).
\end{equation}

Let us expand $\widetilde{\psi}_k(z)$ into Taylor series
\begin{equation}\label{Tayl}
\widetilde{\psi}_k(z)=\sum\limits_{l=0}^{\infty}\, \widetilde{\psi}_{lk}(z-a_k)^l
\end{equation}
in order to sum up $W_{m_1,m_2,k} \widetilde{\psi}_k(z)$
over all translations $m_1+\imath m_2$.

The series $\sum\limits_{j}\, W_{j,k}\widetilde{\psi}_k(z)$, where $j=(m_1,m_2)$ and
$k$ is a fixed number, can be represented via the elliptic Eisenstein functions $E_l(z)$ of order $l$ (see \cite{Weil}):
\begin{equation}\label{Eiz}
\sum\limits_{j}\, W_{j,k}\widetilde{\psi}_k(z)=\sum\limits_{l=0}^{\infty}\,
\overline{\widetilde{\psi}_{lk}}r_k^{2(l+1)}E_{l+2}(z-a_k).
\end{equation}

The series $\sum\limits_{j}{}^{'}\,W_{j,k}\widetilde{\psi}_k(z):=\sum\limits_{j}\,W_{j,k}\widetilde{\psi}_k(z)-
\left(\frac{r_k}{z-a_k}\right)^2\overline{\widetilde{\psi}_k\left(\frac{r_k^2}{\overline{z-a_k}}+a_k\right)}$ can be written in the form
\begin{equation}\label{Mod_Eiz}
\sum\limits_{j}{}^{'}\, W_{j,k}\widetilde{\psi}_k(z)=\sum\limits_{l=0}^{\infty}\,
\overline{\widetilde{\psi}_{lk}}r_k^{2(l+1)}\sigma_{l+2}(z-a_k),
\end{equation}
where $\sigma_l$ is the modified Eisenstein function
defined by the formula $\sigma_l(z):=E_l(z)-z^{-l}$.
The Eisenstein functions $E_l$ converges absolutely and uniformly
for $l=3,4,\dots$ and conditionally for $l=2$ (\cite{Weil}).

Thus, we can rewrite the equations (\ref{sis1}) and (\ref{eq1}) for $\widetilde{\psi}_k$ and $\widetilde{\psi}$ as follows:
\begin{equation}\label{psi_k}
\widetilde{\psi}_k(z)=\sum\limits_{m \neq k}^{N}\sum\limits_{l=0}^{\infty}\,\rho_m \overline{\widetilde{\psi}_{lm}}\,r_m^{2(l+1)}\,E_{l+2}(z-a_m)+
\sum\limits_{l=0}^{\infty}\,\rho_k\overline{\widetilde{\psi}_{lk}}\,r_k^{2(l+1)}\,\sigma_{l+2}(z-a_k)+1,
\end{equation}
\begin{equation}\label{psi}
\widetilde{\psi}(z)=\sum\limits_{m =1}^{N}\sum\limits_{l=0}^{\infty}\,\rho_m\overline{\widetilde{\psi}_{lm}}\,r_m^{2(l+1)}\,E_{l+2}(z-a_m).
\end{equation}

Now we need to find the numerical coefficients $\widetilde{\psi}_{lm}$ of the system (\ref{psi_k}). Note that the equation (\ref{psi}) for $\widetilde{\psi}$ has the same
coefficients $\widetilde{\psi}_{lm}$. Taking a partial sum of Taylor series with $M$ first items
$$
\widetilde{\psi}_k(z)=\widetilde{\psi}_{0k}+\widetilde{\psi}_{1k}(z-a_k)+\widetilde{\psi}_{2k}(z-a_k)^2+\dots+\widetilde{\psi}_{Mk}(z-a_k)^M
$$
and collecting the coefficients of the like powers of $z-a_k$,
we obtain the formula for definition of $\widetilde{\psi}_{jk}$:
\begin{equation}
\widetilde{\psi}_{jk}=\frac{1}{j!} \widetilde{\psi}^{(j)}_k \Big|_{z=a_k},
\end{equation}
where $\widetilde{\psi}^{(j)}_k$ is derivative of order $j$ of the function $\widetilde{\psi}_k$. Then, we get
$$
\widetilde{\psi}_{jk}=\frac{1}{j!}\sum\limits_{m \neq k}^{N}\sum\limits_{l=0}^{M}\,\rho_m\overline{\widetilde{\psi}_{lm}} r_m^{2(l+1)} (-1)^j \frac{(l+j+1)!}{(l+1)!} E_{l+j+2}(a_k-a_m)
$$
\begin{equation} \label{psi_jk}
+\frac{1}{j!}\sum\limits_{l=0}^{M}\,\rho_k\overline{\widetilde{\psi}_{lk}} r_k^{2(l+1)} (-1)^j \frac{(l+j+1)!}{(l+1)!} \sigma_{l+j+2}(0)+I_j,
\end{equation}
where $I_j=\left\{
\begin{array}{lc}
1,  \; j=0, \\
0, \; j=1,\dots,M.
\end{array}
\right.$

Thus, we arrive at the system with $N(M+1)$ unknown constants $\widetilde{\psi}_{jk}$ and $N(M+1)$ equations which can be solved numerically. Note that this system is obtained for an arbitrary number $N$ of inclusions.

\begin{remark}\label{rem:I21}
 Note that doubly periodicity of $E_{l+2}$ and the relations $E'_{l+2}=-(l+2)E_{l+3}$, $l=0,1,2,\dots$ imply $\int\limits_{-0.5}^{0.5}\, E_{l+2}(x+0.5\imath)\,dx=0$ for $l=1,2,\dots$, while the equality $\int\limits_{-0.5}^{0.5}\, E_{2}(x+0.5\imath)\,dx=0$ can be obtained by numerical calculation. Therefore,  $\int\limits_{-0.5}^{0.5}\, \widetilde{\psi}(x+0.5\imath)\,dx=0$.
\end{remark}

\begin{remark}
Note that for finding of the flux distribution (namely, the functions $\widetilde{\psi},\widetilde{\psi}_k,\dots,\widetilde{\psi}_N$) in an explicit form, we change an algorithm of solution of the equations (\ref{psi_k}) and (\ref{psi}) in comparison with the algorithm represented in \cite{BeMi01}. It allows to get more accurate numerical values of the flux in each point of considered composite material.
\end{remark}

\begin{remark}
Note that it is possible to find the heat flux distribution in each cell $Q_{(m_1,m_2)}$ with
corresponding centers $a_1+m_1+\imath m_2$, $a_2+m_1+\imath m_2$, $a_3+m_1+\imath m_2$, $a_4+m_1+\imath m_2$.
\end{remark}

Thus in the next section when discussing numerical results we are concentrating only on the computations in the $Q_{(0,0)}$ unit cell.
%%%%%%%%%%%%%%%%%%%%%%%%%%%%%%%%%%%%%%%%%%%%%%%%%%%%%%%%%%%%%%%%%%%%%

\section{Numerical results and discussions}

\subsection{The flux and the temperature distribution}

First we indirectly check the performance of the modified
algorithm described in (\ref{psi_k}) -- (\ref{psi_jk}).
As an example, we consider the case when four inclusions are situated within one cell, i.e., $N=4$.
We suppose throughout the computations that the heat flux of the fixed intensity $A=-1$ flows in different directions with respect to the main axis. Here the minus sign shows that the flux is directed from the right to the left (or from the top to the bottom) depending on the angle $\theta$. The conductivity of the matrix is set as $\lambda_m=1$, while those for the inclusions,
$\lambda_k$, will take different values. The algorithm is realized in Maple 14 software.

We take for the first test a non-symmetrical configuration, with respect to $Ox$-axis,
with two inclusions from the neighboring cells are situated very close to each other
as depicted on Figure~\ref{cell_00}. The centers of the inclusions are situated in the points:
\begin{equation}
\label{centers}
a_1=-0.18+0.2 \imath,\quad a_2= 0.33-0.34\imath, \quad a_3=0.33+0.35 \imath, \quad a_4= -0.18-0.2\imath,
\end{equation}
while their radii are the same $r_k=R$.

\begin{figure}
\begin{center}
\resizebox*{5cm}{!}{\includegraphics{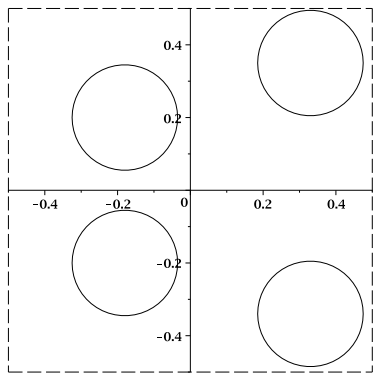}}%
\caption{The cell $Q_{(0,0)}$.}%
\label{cell_00}
\end{center}
\end{figure}

Convergence of the results computed for various numbers of the truncation parameter, $M$, showing how many terms are selected for computations in the Taylor series (\ref{Tayl}) is analyzed in the Table~\ref{table2}.

The flux components $Q_x$ and $Q_y$ in the center $a_k$ of $k$-inclusion
$$
Q^{(k)}_x(a_k)\equiv\lambda_k \frac{\partial T_k(a_k)}{\partial x}=\frac{2 \lambda_k \lambda_m}{\lambda_m+\lambda_k} \cdot{\mathrm{Re}}\,\psi_k(a_k),
$$
$$
Q^{(k)}_y(a_k)\equiv\lambda_k \frac{\partial T_k(a_k)}{\partial y}=- \frac{2 \lambda_k \lambda_m}{\lambda_m+\lambda_k}\cdot {\mathrm{Im}}\,\psi_k(a_k),
$$
are calculated in accordance with the formula (\ref{Temp}). The flux components in any point of the matrix can be found as
$$
Q^{(m)}_x(z)=\lambda_m  \cdot{\mathrm{Re}}\,(\psi(z)+B),\,\quad
Q^{(m)}_y(z)=-\lambda_m  \cdot{\mathrm{Im}}\,(\psi(z)+B).
$$
We calculate $Q^{(m)}_x(z)$ and $Q^{(m)}_y(z)$ at the point $z=0\in D_{matrix}$
belonging to the matrix. Computations in the Table~\ref{table2} are given for four consequent value of the truncation number $M$ ($M=0,1,...,4)$, and fixed other parameters: $\theta=0$, $R=0.145$,
$\lambda_m=1$, $\lambda_k=100$.

\begin{table}
\tbl{The flux components for different truncation number $M$, while other problem parameters are: $\theta=0$, $R=0.145$, $\lambda_m=1$, $\lambda_k=100$ and the configurations of inclusion defined by (\ref{centers}).}
  {\begin{tabular}{@{}ccccc}\toprule
        $M$ & $Q^{(1)}_x(a_1)$ & $Q^{(1)}_y(a_1)$ & $Q^{(m)}_x(0)$ & $Q^{(m)}_y(0)$  \\
  \colrule
     $0$ & $1.50390574$ & $-0.00595137$ & $0.75430758$ & $-0.00150794$  \\
     $1$ & $1.51410894$ & $-0.00549234$ & $0.74026814$ & $-0.00170226$  \\
     $2$ & $1.51491054$ & $-0.00568751$ & $0.73098974$ & $-0.00176024$ \\
     $3$ & $1.51472013$ & $-0.00569043$ & $0.73081787$ & $-0.00176227$  \\
     $4$ & $1.51472348$ & $-0.00569169$ & $0.73081950$ & $-0.00176386$  \\
     \botrule
  \end{tabular}}
\label{table2}
\end{table}
Computations performed for this configuration and the material parameters suggest that, taking $M=4$, the accuracy is between five or six valid units depending on where the flux is computed.
Note that the material contrast is rather high thus the accuracy is good enough inside both the materials (in the matrix and in the inclusions).
Clearly, the value of the inclusion radius affects the accuracy of the computations. Moreover, for
symmetric configurations of the inclusions the accuracy is higher.

As an example, we also show the flux distribution inside the cell $Q_{(0,0)}$ for different angles and conductivities of inclusions on Figure~\ref{N4_la0.01_Teta45}-\ref{N4_la100_Teta45}. We take for calculations the centers of inclusions in the points (\ref{centers}) and the radius $R=0.145$.
\begin{figure}
\begin{center}
\subfigure{\resizebox*{6cm}{!}{\includegraphics{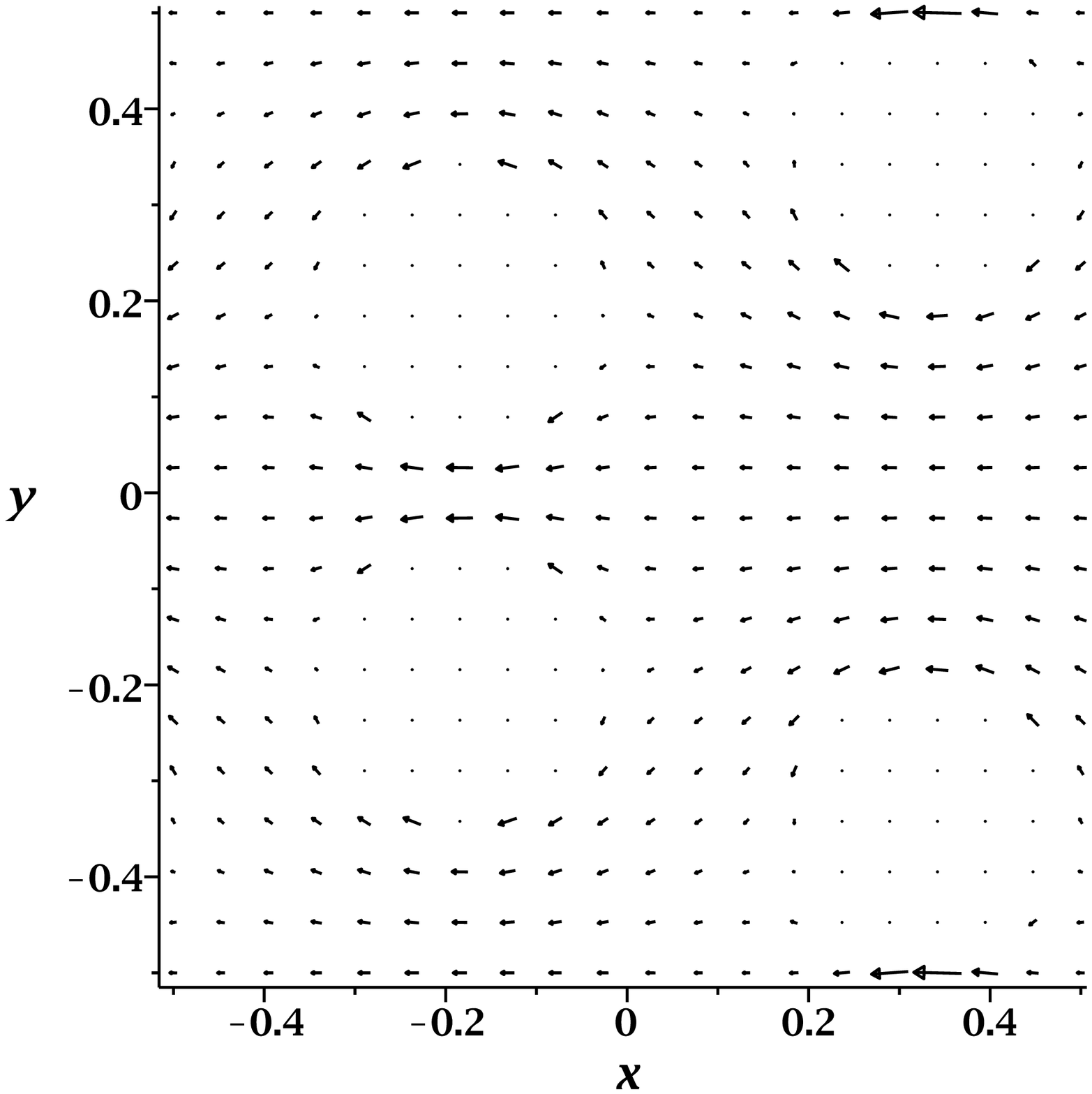}}}%
\subfigure{\resizebox*{6cm}{!}{\includegraphics{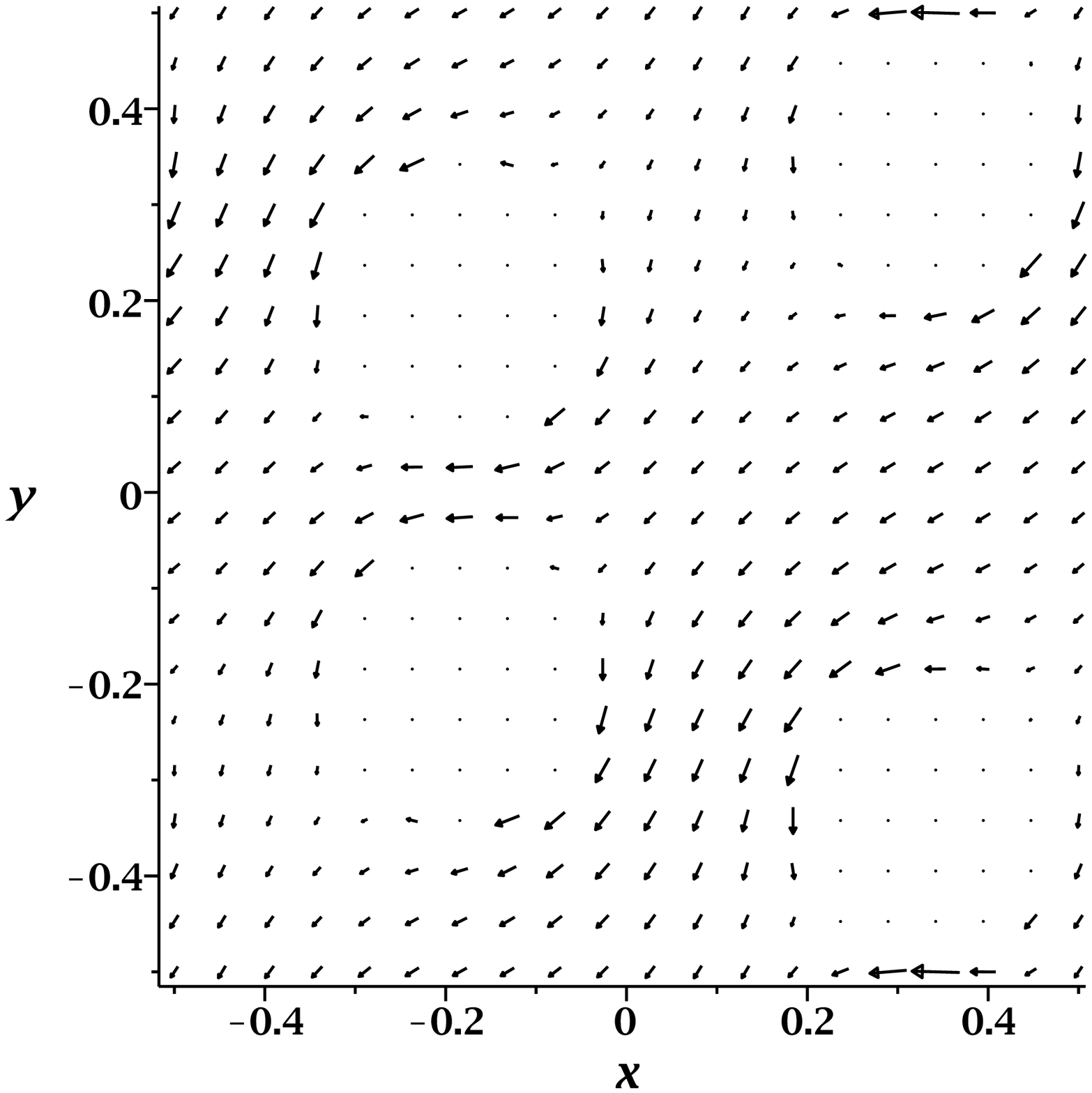}}}%
\caption{The flux distribution inside $Q_{(0,0)}$ for $\lambda_k=0.01$, $\theta=0; \pi/4$.}%
\label{N4_la0.01_Teta45}
\end{center}
\end{figure}

\begin{figure}
\begin{center}
\subfigure{\resizebox*{6cm}{!}{\includegraphics{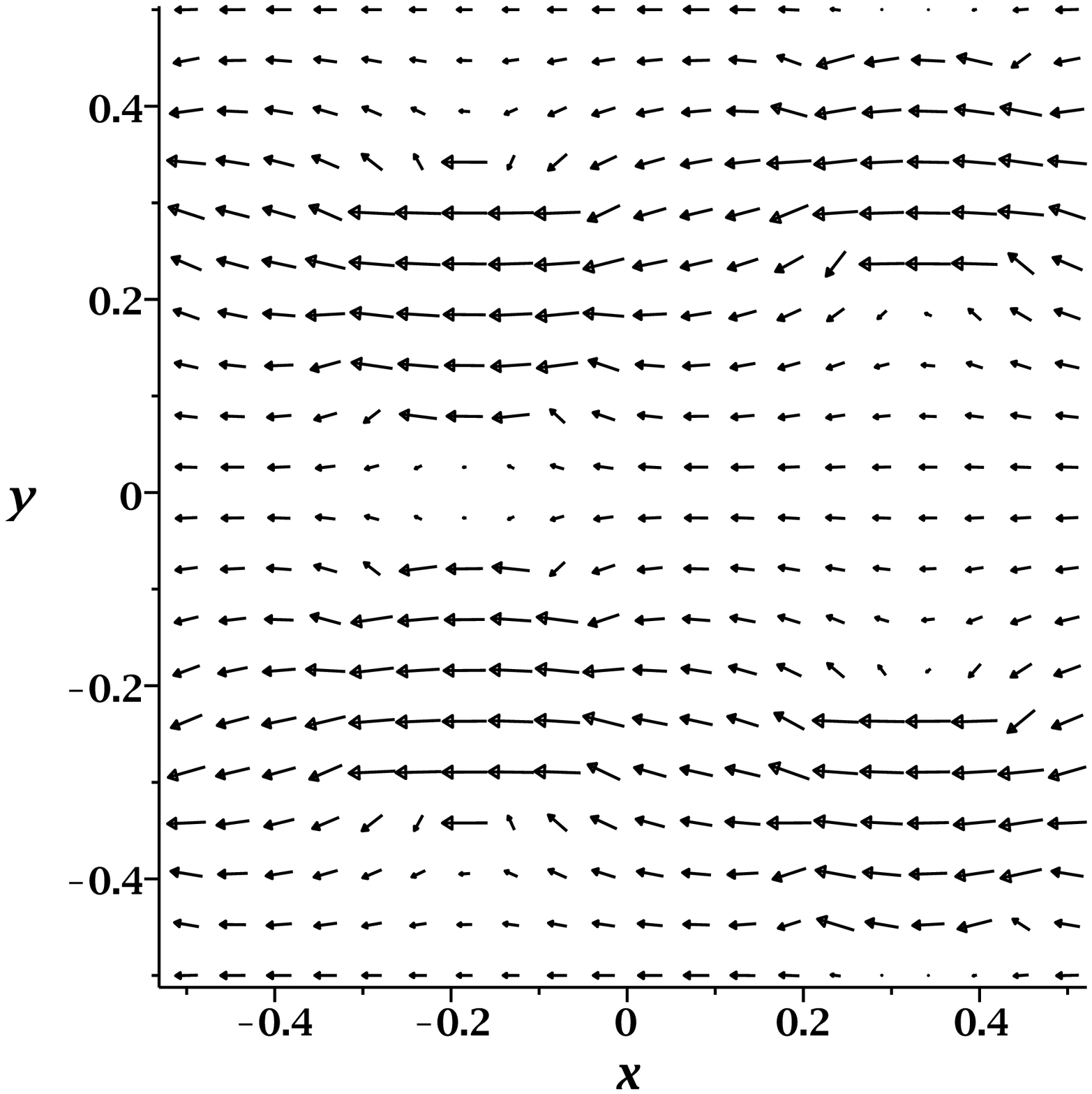}}}%
\subfigure{\resizebox*{6cm}{!}{\includegraphics{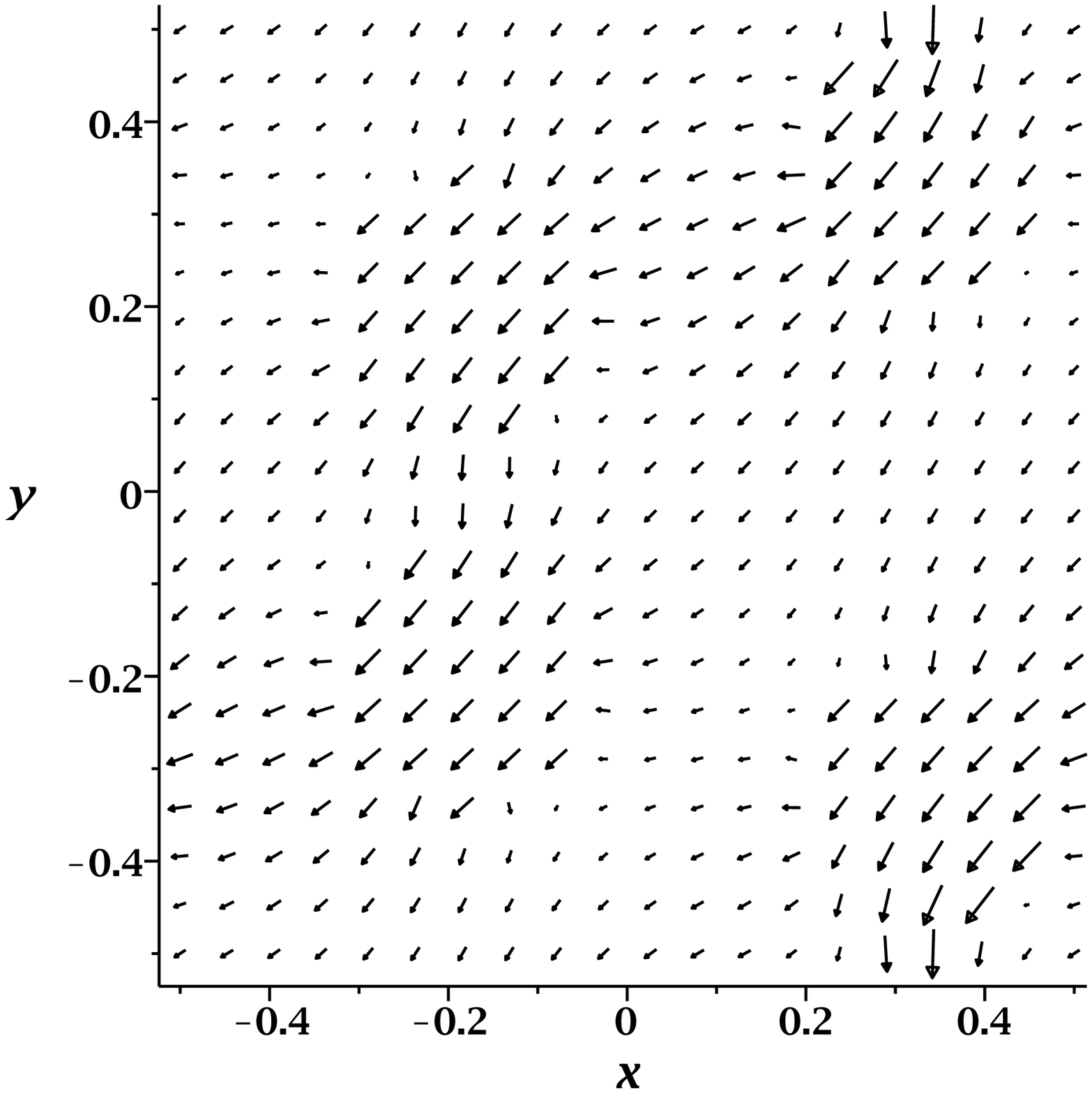}}}%
\caption{The flux distribution inside $Q_{(0,0)}$ for $\lambda_k=100$, $\theta=0; \pi/4$.}%
\label{N4_la100_Teta45}
\end{center}
\end{figure}

According to (\ref{potential_1}) in order to find the temperature function $T=T(x,y)$, one needs to find functions $\varphi$ and $\varphi_k, \, k=1,\dots,N$. Using (\ref{flux_funs}) one can recover it, up to arbitrary constant, integrating functions $\psi$ and $\psi_k,\, k=1,\dots,N$. To define the constant, we use the boundary conditions (\ref{10}). As a result of the integration, Weierstrass zeta-function appears  (cf. Appendix A). The temperature distribution $T(x,y)$ is presented on Figure~\ref{Temp_distr1}-\ref{Temp_distr2}. The same two different contrasts ratios ($0.01$ and $100$) as for the flux distributions are given.
Namely, we consider the following set of the parameters: $\lambda_m=1$, $R=0.145$, $\theta=0; \frac{\pi}{4}$ with $\lambda_k=100$ and $\lambda_k=0.01$.
\begin{figure}
\begin{center}
\subfigure{\resizebox*{6cm}{!}{\includegraphics{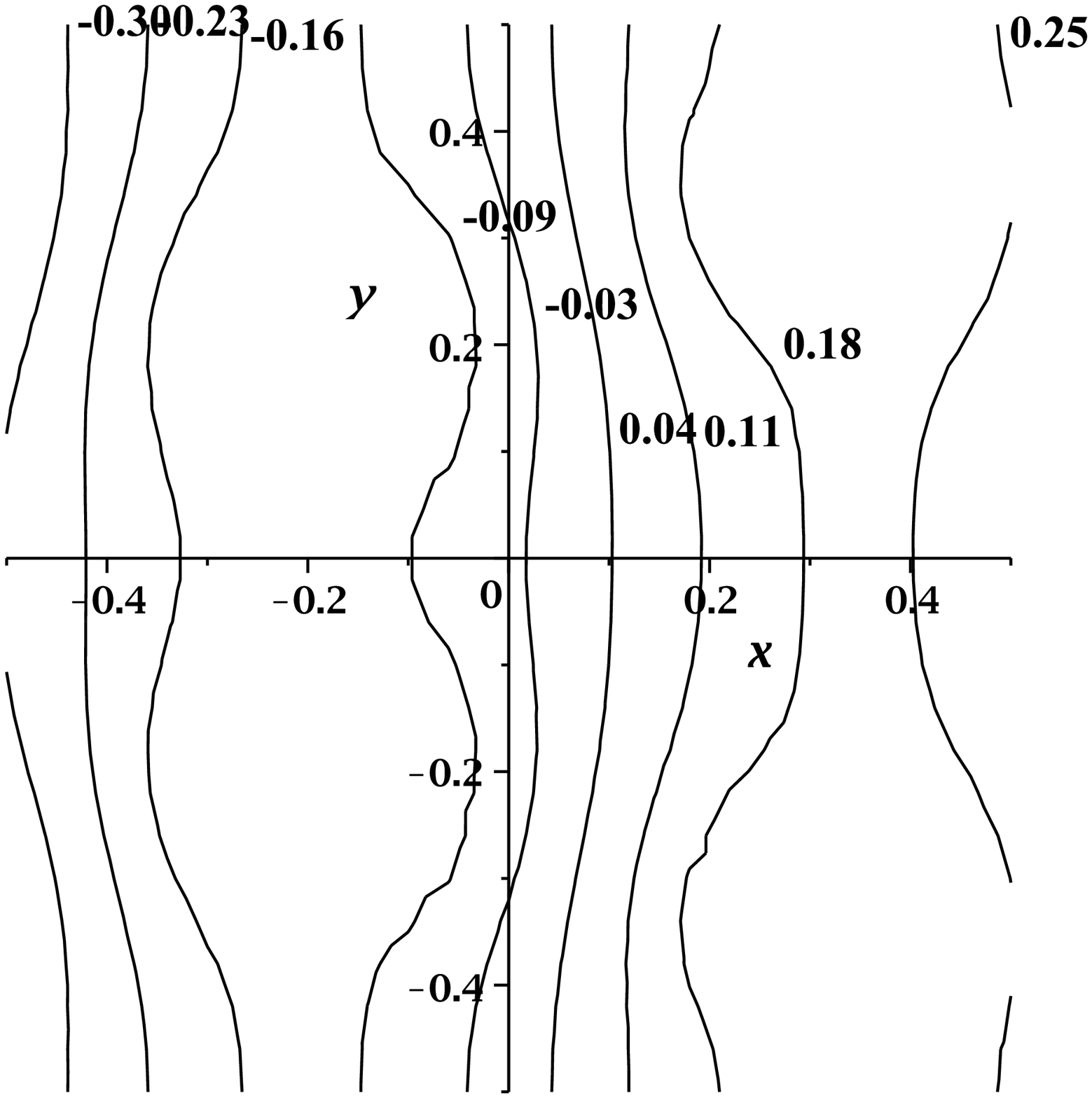}}}%
\subfigure{\resizebox*{6cm}{!}{\includegraphics{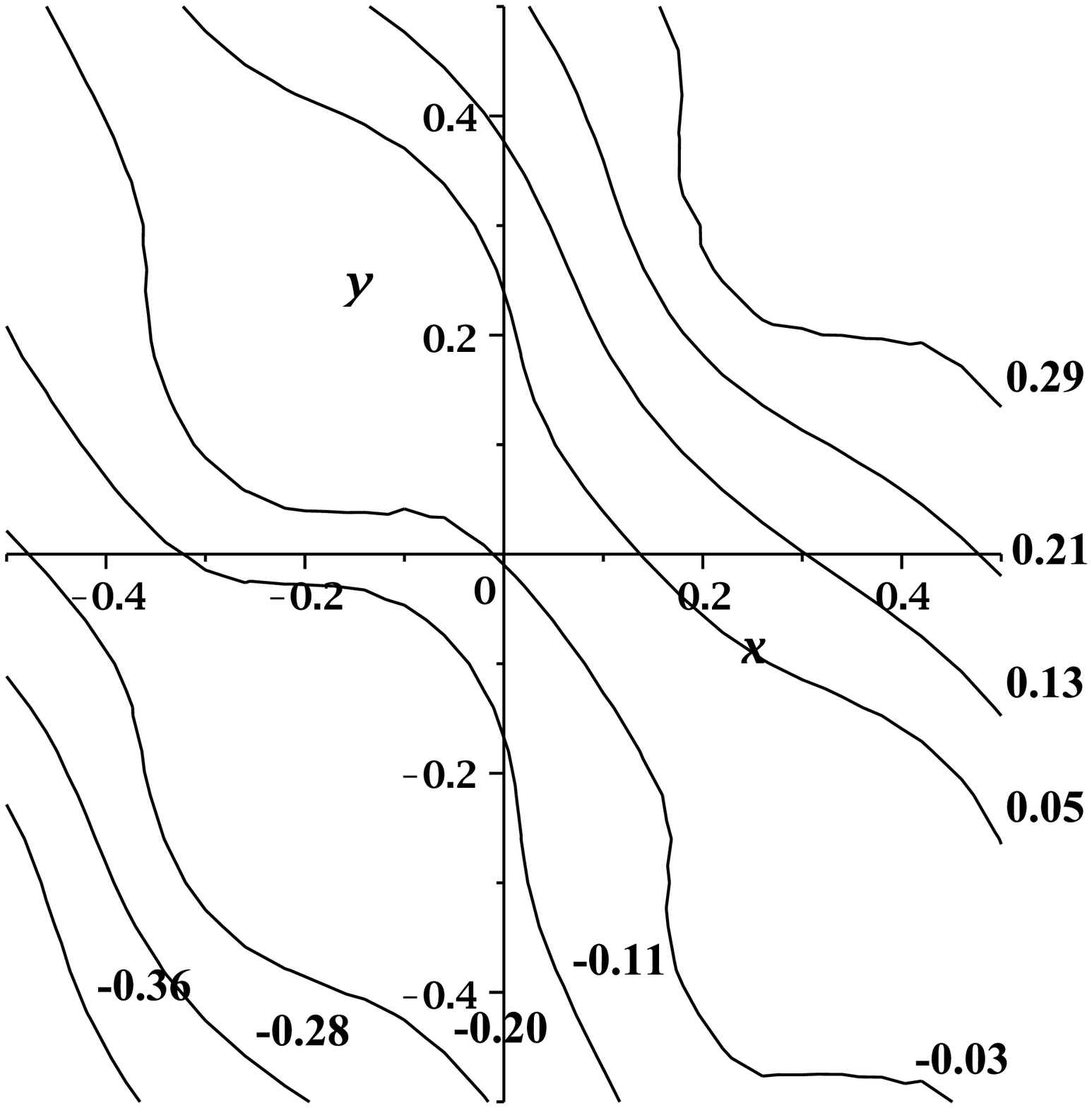}}}%
\caption{The temperature distribution inside $Q_{(0,0)}$ for $\lambda_k=100$, $\theta=0; \pi/4$.}%
\label{Temp_distr1}
\end{center}
\end{figure}

\begin{figure}
\begin{center}
\subfigure{\resizebox*{6cm}{!}{\includegraphics{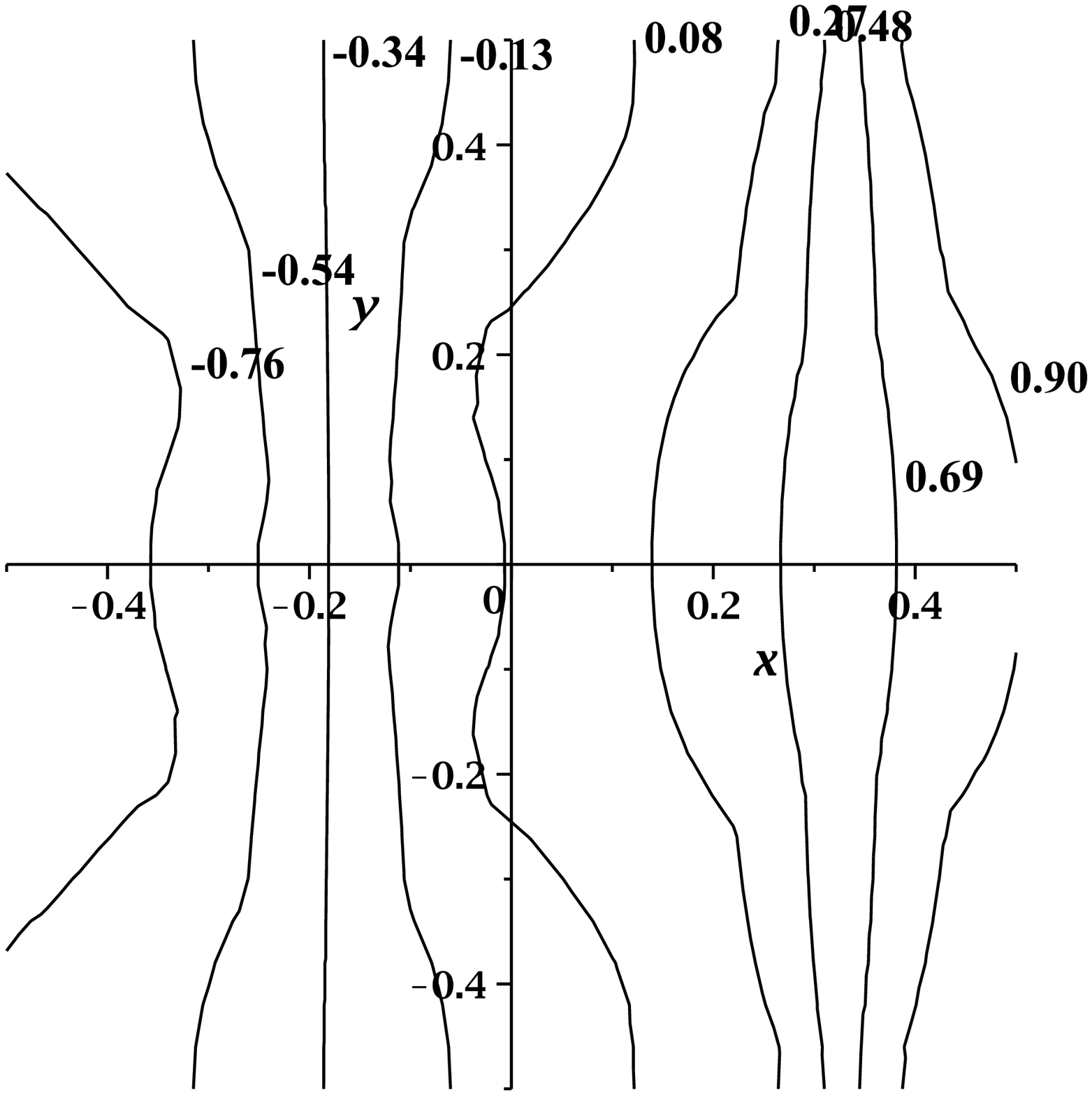}}}%
\subfigure{\resizebox*{6cm}{!}{\includegraphics{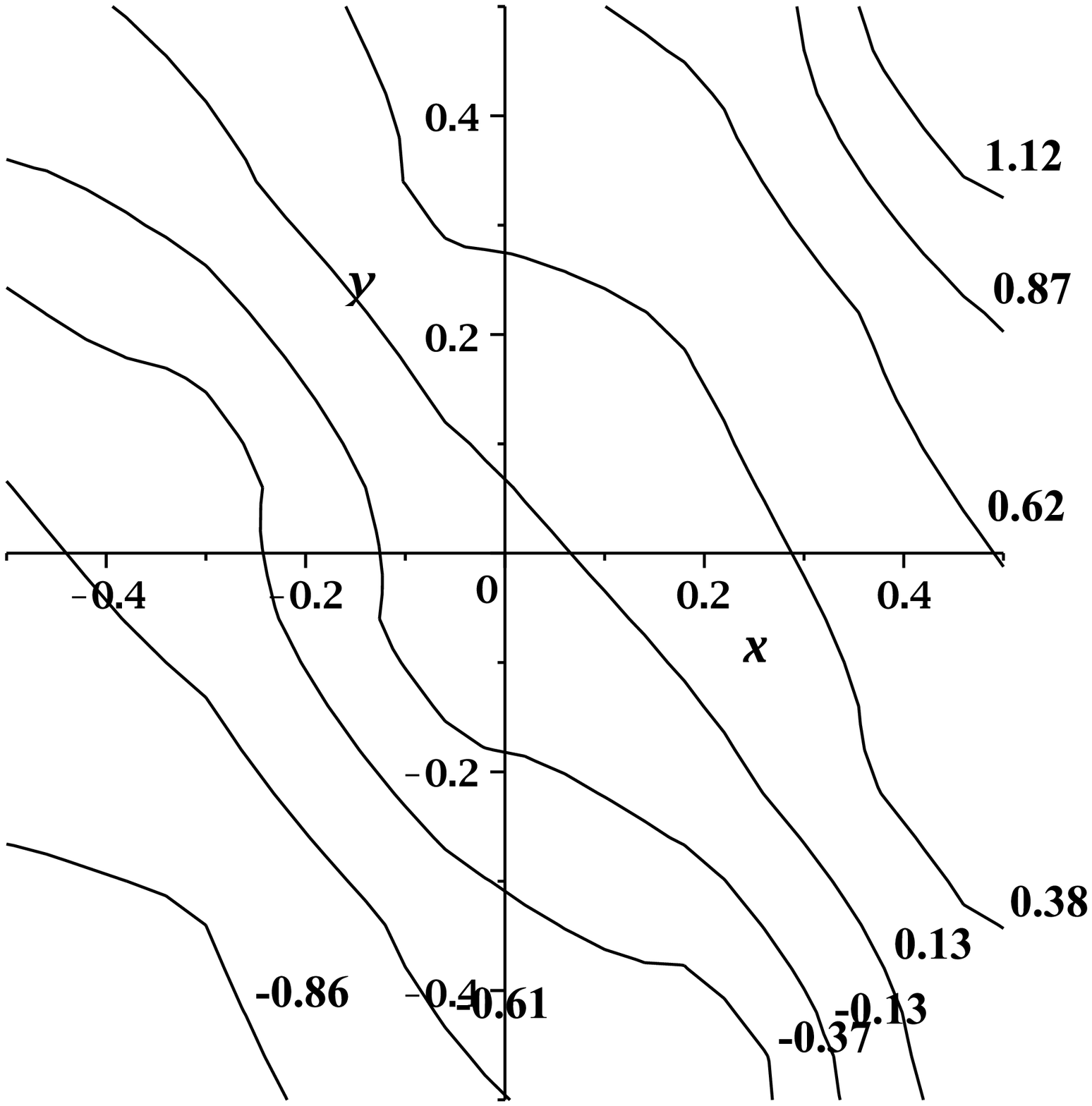}}}%
\caption{The temperature distribution inside $Q_{(0,0)}$ for $\lambda_k=0.01$, $\theta=0; \frac{\pi}{4}$.}%
\label{Temp_distr2}
\end{center}
\end{figure}

\subsection{Average properties of the composite}

The effective conductivity of microscopically isotropic composite materials with the inclusion volume fraction:
\[
\nu=\sum_{k=1}^N \pi r_k^2\ll1
\]
is defined by the spherical tensor $\Lambda_e=\lambda_e {\bf I}$ where the effective conductivity
is computed by the classical Maxwell formula
\begin{equation}\label{LrelCM}
\lambda_e=\frac{1+\rho \nu}{1-\rho \nu}+\mathcal{O}(\nu^2),\quad \nu\to0,
\end{equation}
where $\lambda_m=1$, $\rho=(\lambda_k-1)/(\lambda_k+1)$.
For the history and applications of this formula see for example \cite{Mi00}.

To compare the results obtained on the base of the improved algorithm and the classical Maxwell formula, we choose a periodic array of four inclusions ($N=4$) placed symmetrically in the unit cell with the centers: $a_1=-0.25+0.25 \imath$, $a_2= 0.25+0.25\imath$, $a_3=0.25-0.25 \imath$, $a_4=-0.25-0.25\imath$. Other parameters are: the flux flows along $Ox$-axis ($\theta=0$), $A=-1$, inclusions conductivities $\lambda_k=100$ and various values of the concentration parameter $\nu$. In Table~\ref{table3}, a comparison of results calculated by two methods is presented. Here the value $\delta \lambda=(\lambda_e-\lambda_e^M)/\lambda_e$ indicates the accuracy of the approximate formula. We also supplement the computations by the results
obtained using the FORTRAN code given in \cite{Kusch}.
\begin{table}
\tbl{Comparison of the effective conductivities: $\lambda_e$, $\lambda_e^{M}$ and $\lambda_e^*$ computed by the new algorithm,
classic Maxwell formula  and the algorithm presented in \cite{Kusch} for the symmetrically situated inclusions when
$\lambda_m=1, \lambda_k=100$, $M=4$.}
 {\begin{tabular}{@{}cccccccccc}\toprule
      $R$ &$\nu$ &$D/d$& $\lambda_e^{M}$  & $\lambda_e$ &$\delta \lambda$ & $\frac{\lambda_e-\lambda_e^{M}}{\nu^2}$&  $\lambda_e^*$\\
    \colrule
     $0$ & $0$ &$0$& $1$ & $1$ & 0 &$-$&$1$\\
     $0.005$ & $0.00031$ & $0.0204$ & $1.00062$ & $1.00062$ & $6.8 \cdot 10^{-19}$ & $6.9\cdot 10^{-12} $& $1.00062$ \\
     $0.01$ & $0.00126$ & $0.0417$ & $1.00247$ & $1.00247$ & $1.8 \cdot 10^{-15}$& $1.1 \cdot 10^{-9}$ & $1.00247$  \\
     $0.03$ & $0.01131$ & $0.1364$ & $1.02242$ & $1.02242$  & $1.1 \cdot 10^{-10}$& $8.5 \cdot 10^{-7}$&  $1.02242$  \\
     $0.07$ & $0.06158$ & $0.3889$ &  $1.12847$ & $1.12847$  & $5.1 \cdot 10^{-7} $& $1.5 \cdot 10^{-4}$&  $1.12848$  \\
     $0.1$ & $0.12566$ & $0.6667$ & $1.28096$ &   $1.28098$  & $1.8 \cdot 10^{-5}$& $1.5 \cdot 10^{-3}$&  $1.28097 $ \\
     $0.12$ & $0.18095$ & $0.9231$ & $1.43123$ &  $1.43140$ & $1.6 \cdot 10^{-4}$& $5.0 \cdot 10^{-3}$&  $1.43138$  \\
     $0.145$ & $0.26421$ & $1.3810$ & $1.69897$ &  $1.70032$ & $8.0 \cdot 10^{-4}$& $1.9 \cdot 10^{-2}$&  $1.70033$  \\
     $0.185$ & $0.43008$ & $2.8462$ & $2.45762$ &   $2.48345$ & $0.0104$& $0.1397$&  $2.48342$  \\
     $0.2$ & $0.50265$ & $4.0$ & $2.94245$ &  $3.01755$ & $0.0249$& $0.2972$&  $3.01742$  \\
  \botrule
  \end{tabular}}
\label{table3}
\end{table}

\begin{table}
\tbl{Comparison of the effective conductivities $\lambda_e$ computed by the new algorithm and the classic
Maxwell formula $\lambda_e^{M}$ ($\lambda_m=1, \lambda_k=1/100$).}
 {\begin{tabular}{@{}ccccccc}\toprule
        $R$ &$\nu$ &$D/d$& $\lambda_e^{M}$  & $\lambda_e$ & $\delta \lambda$ & $\frac{\lambda_e-\lambda_e^{M}}{\nu^2}$\\
    \colrule
     $0$ & $0$ &$0$& $1$ & $1$ & 0 & $-$\\
     $0.005$ & $0.00031$ & $0.0204$ & $0.99938$ & $0.99938$ &$-2.4 \cdot 10^{-18}$& $-2.9 \cdot 10^{-11}$\\
     $0.01$ & $0.00126$ & $0.0417$ & $0.99754$ & $0.99754$  &$-1.8 \cdot 10^{-15}$& $-1.2 \cdot 10^{-9}$\\
     $0.03$ & $0.01131$ & $0.1364$ & $0.97807$ & $0.97807$ & $-1.1 \cdot 10^{-10}$& $-8.2 \cdot 10^{-7}$\\
     $0.07$ & $0.06158$ & $0.3889$ & $0.88616$ & $0.88616$  &$-5.1 \cdot 10^{-7}$& $-1.2 \cdot 10^{-4}$\\
     $0.1$ & $0.12566$ & $0.6667$ &  $0.78067$ & $0.78065$  &$-1.8 \cdot 10^{-5}$& $-9.1 \cdot 10^{-4}$\\
     $0.12$ & $0.18095$ & $0.9231$ & $0.69870$ & $0.69862$ & $-1.6 \cdot 10^{-4}$& $-2.5 \cdot 10^{-3}$\\
     $0.145$ & $0.26421$ & $1.38095$&$0.58859$ & $0.58812$ & $-8.0 \cdot 10^{-4}$& $-6.7 \cdot 10^{-3}$\\
     $0.185$ & $0.43008$ & $2.8462$ &$0.40670$ & $0.40267$ & $-0.0105$& $-0.0229$\\
     $0.2$ & $0.50265$ & $4.0$ &     $0.33985$ & $0.33139$ & $-0.0255$& $-0.0335$\\
    \botrule
  \end{tabular}}
\label{table4}
\end{table}

It is not a surprise that the results of the computations done for small inclusion concentrations ($R=0.005;\, 0.01;\,0.03$)
exhibit much better accuracy than that discussed in the Table~\ref{table2}.
Here, $D=2R$, $d$ is a distance between inclusions giving an impression how far (close) to each other they are.
On the contrary, the last two values of the volume fraction $\nu$ stand for rather high inclusion concentration. Our results confirm those obtained earlier
(see, for example, \cite{Kusch,Kach_Sev}) that the Maxwell formula provide a very good accuracy for a regular distribution of the
inclusions up to the level of $30\%$ in the volume fraction. As follows from the computations, to have the deviation between the
formulas less than $1\%$, the distance between the inclusions should be larger than their diameters at least two times.
Clearly, the results refer to the particular chosen configuration of inclusions, but interestingly does not much depend on the contrast parameter $\rho$.

From the results presented in the tables one could try to estimate the constants in the reminder ${\cal O}(\nu^2)$ from (\ref{LrelCM}). However, the reminder in this particular case is rather of the order
${\cal O}(\nu^4)$. Clearly this should relate to the specific configuration of the inclusions in the composite. The authors failed to prove this or to find any known result supporting such a statement.

Finally, we estimate the accuracy of our computations by direct comparison
 with other known benchmarks. In the \cite{McP}, the simplest geometry, one
central inclusion was considered as one of the computational examples. Here, we present the respective date in the Table~\ref{table5}. In parentheses we also indicate the number of the terms in the Taylor series chosen by the authors to guarantee the desirable accuracy. We also compare our computations performed for $M=4$ with those computed for the same number of iterations by the FORTRAN programme from \cite{Kusch}. For an additional check, we also present in the table the results computed for four symmetrical inclusions (which topologically and mechanically equivalent to the only central inclusions with the same volume fracture preserved). All the results are in a perfect agreement (taking into account the number of terms used in the computations of the Taylor series). Surprisingly, increasing number of the inclusions within the unit cell (preserving the composite topology), we achieved a slightly better results.

\begin{table}
\tbl{The effective conductivities $\lambda_e$ computed by the new algorithm (for $M=4$),
the Perrins formula $\lambda_e^{P}$ (\cite{McP}) (number of the terms is given in the parentheses), the classic
Maxwell formula $\lambda_e^{M}$ and the algorithm from \cite{Kusch} ($M=5$) for $\lambda_m=1, \; \lambda_k=50$.}
  {\begin{tabular}{@{}ccccccc}\toprule
         $\nu$ &$\lambda_e \;(N=1)$&$\lambda_e \;(N=4)$ & $\lambda_e^P\,(M)$ & $\lambda_e^M$& $\lambda_e^*$ \\
    \colrule
     $0.1$ & $1.21259$ & $1.21259$ &$1.2126\,(1)$&    $1.21258$ & $1.21259$ \\
     $0.2$ & $1.47599$ & $1.47599$& $1.4760\,(2)$ &   $1.47573$ & $1.47599$ \\
     $0.3$ & $1.81253$ & $1.81253$& $1.8125\,(2)$ &   $1.80992$&  $1.81253$ \\
     $0.4$ & $2.26325$ & $2.26325$& $2.2633\,(4)$ &   $2.24841$&  $2.26325$ \\
     $0.5$ & $2.91440$ & $2.91447$& $2.9146\,(4)$ &   $2.84906$&  $2.91440$ \\
     $0.55$ &$3.37256$  &$3.37250$& $3.3732\,(-)$ &   $3.24116$&  $3.37256$ \\
     $0.6$ & $3.98590$ & $3.98653$& $3.9881\,(8)$&    $3.72222$&  $3.98590$ \\
     \botrule
  \end{tabular}}
\label{table5}
\end{table}

Now we return back to the original asymmetric configuration (\ref{centers}).
The components $\lambda_e^{ij}$ are calculated using the formula given in the Appendix B with use of the new algorithm.
We present the computations for the materials conductivities  $\lambda_m=1$ and $\lambda_k=100;\,1/100$.
Values of all components of the tensor $\Lambda_e$ as a function on the concentration $\nu$ are presented in Table~\ref{table6} and in Table~\ref{table7}.

\begin{table}
\tbl{The components of the effective conductivity tensor $\Lambda_e$ for the configuration of the inclusions given in (\ref{centers})
for the material constants $\lambda_k=100$, $\lambda_m=1, M=4$.}
  {\begin{tabular}{@{}cccccccc}\toprule
     $R$ & $\nu$ & $\lambda_e^x$  & $\lambda_e^{yx}$ & $\lambda_e^y$  & $\lambda_e^{xy}$  \\
  \colrule
     $0$ & $0$  & $1$ & $0$ & $1$ & $0$ \\
     $0.05$& $0.03142$ & $1.06285224$ & $2.434 \cdot 10^{-7}$ & $1.06425870$ & $2.434\cdot 10^{-7}$ \\
     $0.11$& $0.15205$ & $1.331641$ & $0.000008$ & $1.375325$ & $0.000008$ \\
     $0.135$& $0.22902$ & $1.53413$ & $0.000027$ & $1.6630$ & $0.000027$  \\
     $0.145$& $0.26421$& $1.6381$ & $0.000047$ & $1.844$ & $0.000047$\\
  \botrule
  \end{tabular}}
\label{table6}
\end{table}

\begin{table}
\tbl{The components of the effective conductivity tensor $\Lambda_e$ for the configuration of the inclusions given in (\ref{centers}) for the material constants $\lambda_k=1/100$ and $\lambda_m=1$.}
  {\begin{tabular}{@{}cccccccc}\toprule
      $R$ & $\nu$ & $\lambda_e^x$  & $\lambda_e^{yx}$ & $\lambda_e^y$  & $\lambda_e^{xy}$  \\
    \colrule
     $0$& $0$  & $1$ & $0$ & $1$ & $0$ \\
     $0.05$& $0.03142$ & $0.93962116$ & $2.152 \cdot 10^{-7}$ & $0.94086455$ & $2.152 \cdot 10^{-7}$ \\
     $0.11$& $0.15205$ & $0.727102$ & $0.00000443$ & $0.750954$ & $0.00000443$  \\
     $0.135$& $0.22902$ & $0.60148$ & $0.000011$ & $0.65186$ & $0.000011$  \\
     $0.145$& $0.26421$& $0.5433$ & $0.000016$ & $0.611$ & $0.000016$ \\
    \botrule
  \end{tabular}}
\label{table7}
\end{table}

Thus the composite described by such configuration of the inclusions is not isotropic. The anisotropy increases with the value of the radius $R$ (or, equivalently, with the volume fraction $\nu$). The level of the computational accuracy in the Table~\ref{table6} and Table~\ref{table7} was controlled by a stabilization of the meaningful numbers.

To conclude the paper, we have proposed here the improved algorithm to solve  the system (\ref{sisPhi}).
It allows us to reconstruct the solution and its gradient at an arbitrary point of the composite and the effective properties of the composite.
We have shown an effectiveness of the improved algorithm on several examples and discussed peculiarities of the computations related to a particular configuration of the inclusions. And last but not least, the algorithm constructed in this paper may
be used to compute practically arbitrary configurations of the inclusions in the composite using the benefits of the functional equation approach.

\section*{Acknowledgements}
D. Kapanadze and E. Pesetskaya are supported by Shota Rustaveli National Science
 Foundation within grant FR/6/5-101/12 with the number 31/39. G. Mishuris is grateful
to the FP7 PEOPLE IRSES Project TAMER under number 610547 for support of this research.
%%%%%%%%%%%%%%%%%%%%%%%%%%%%%%%%%%%%%%%%%%%%%%%%%%%%%%%%%%%%%%%%%%%%

\appendices

\section{Description of the Eisenstein function of first order}

This section contains the description of the Eisenstein function $E_1$ of first order.
Properties of the Eisenstein functions of higher orders are described in details in \cite{Weil}.

The theory of elliptic functions provides the following formula for a lattice sum introduced by Rayleigh
$$
S_{2n}:=\sum\limits_{m_1,m_2}{}^{'}\,\frac{1}{(m_1+\imath m_2)^{2n}}.
$$
The Eisenstein functions of order $p$ are defines as
$$
E_p(z):=\sum\limits_{m_1,m_2}\,(z-m_1-\imath m_2)^{-p}.
$$
The function $E_2$  and the Weierstrass function $\wp$ are related by the identities (\cite{Hurv})
$$
E_2(z):=\wp(z)+S_2,
$$
where $S_2=\pi$ for the square array.

The derivative of the Eisenstein function possesses the following property:
$$
E'_p(z)=-p\cdot E_{p+1}(z).
$$
Using this relation, we have $E'_1(z)=-E_2(z)$. Thus, integration on $[0,z]$ gives
$$
E_1(z)=\zeta(z)-\pi z,
$$
where $\zeta$ is the Weierstrass zeta function, and $\zeta'(z)=-\wp(z)$.

\section{Evaluation of the effective conductivity}

In general case of composites with different random non-overlapping inclusions the tensor of effective conductivity $\Lambda_e$ has a form
\begin{equation}
\label{Lambda_e}
\Lambda_e=\left(\begin{array}{cc}
\lambda_e^x & \lambda_e^{xy}\\
\lambda_e^{yx} & \lambda_e^y
\end{array}\right)
\end{equation}
with components which can be found from the well-known equation
\begin{equation}\label{LrelMit}
\langle{\bf q} \rangle=-\Lambda_e \cdot \langle\nabla T\rangle,
\end{equation}
where $\langle{\bf q}\rangle=(\mathfrak{q}_1,\mathfrak{q}_2)$ is the average flux, and $\langle\nabla T\rangle=(T_1,T_2)$ is the average temperature gradient with
\begin{equation}
\label{q_1}
\mathfrak{q}_j =\lambda_m \iint\limits_{D_0}\,\frac{\partial
T}{\partial x_j}\,dx_1dx_2 +\sum\limits_{k=1}^N \lambda_k
\iint\limits_{D_k} \frac{\partial T_k}{\partial x_j}\,
dx_1dx_2,
\end{equation}
and
\begin{equation}
\label{T_1}
T_j =\iint\limits_{D_0}\,\frac{\partial
T}{\partial x_j}\,dx_1dx_2 +\sum\limits_{k=1}^N
\iint\limits_{D_k} \frac{\partial T_k}{\partial x_j}\,
dx_1dx_2,
\end{equation}
where $j=1,2$ and $x_1=x$ and $x_2=y$.

The integrals above can be transformed using first Green's formula
\begin{equation}
\label{Green}
\int_U \left( \psi \Delta \varphi + \nabla \varphi
\cdot \nabla \psi\right)\, dV = \oint_{\partial U} \psi \left(
\nabla \varphi \cdot \bold{n} \right)\, dS
\end{equation}
with $\psi=x$ or $\psi=y$  and $\varphi(x,y)=T$ in $D_0$ (or $\varphi(x,y)=T_k$
in the respective domain $D_k$).
Moreover,
\begin{equation}
\label{lam_x2}
\oint_{\partial D_0} x \frac{\partial
T}{\partial n}ds=\oint_{\partial Q_{(0,0)}} x
\frac{\partial T}{\partial n}ds-\sum\limits_{k=1}^N \oint_{\partial D_k}
x \frac{\partial T_k}{\partial n}ds,
\end{equation}
where the curves $\partial Q_{(0,0)}$ and $\partial D_k$ are oriented in the counterclockwise direction.

The first integral can be directly  computed with use of
({\ref{bound_cell_1}), ({\ref{bound_cell_2}) and (\ref{zero_flux})
\[
 \oint_{\partial Q_{(0,0)}} \hspace{-3mm}x \frac{\partial T}{\partial n}ds=
\int_{-1/2}^{1/2}x(-A\sin\theta)dx-\int_{-1/2}^{1/2}x(-A\sin\theta)dx+
\]
\[
\frac{1}{2}\int_{-1/2}^{1/2}(-A\cos\theta)dy+
\frac{1}{2}\int_{-1/2}^{1/2}(-A\cos\theta)dy=-\frac{A}{\lambda_m}\cos\theta.
\]

Repeating the same line of the reasoning with $\psi=y$ and
$\varphi(x,y)=T$ in $D_0$ (or $\varphi(x,y)=T_k$ in the respective
domain $D_k$) with first Green's formula (\ref{Green})
\begin{equation}
\label{flux_y}
\oint_{\partial Q_{(0,0)}} \hspace{-3mm}y \frac{\partial T}{\partial n}ds=-\frac{A}{\lambda_m}\sin\theta.
\end{equation}

Thus using the Green formula (\ref{Green}), we have
$$
\mathfrak{q}_1=\lambda_m \left(\oint\limits_{\partial Q_{(0,0)}}\, x \frac{\partial T}{\partial n}\, ds -\frac{\lambda_k}{\lambda_m}
\sum\limits_{k=1}^N
\oint\limits_{\partial D_k}x \frac{\partial T_k}{\partial n}\, ds\right)+ \sum\limits_{k=1}^N \lambda_k\oint\limits_{\partial D_k}x \frac{\partial T_k}{\partial n}\, ds.
$$
Using the fact that the contour integrals annihilate each other along the
common boundaries $\partial D_k$ and the formula
({\ref{bound_2}), we finally obtain:
$$
\mathfrak{q}_1=\lambda_m \oint\limits_{\partial Q_{(0,0)}}x \frac{\partial T}{\partial n}\, ds=-A \cos \theta.
$$
Analogously,
$$
\mathfrak{q}_2=-A \sin \theta.
$$
According to the same arguments as above, we have
$$
T_1=-\frac{A \cos \theta}{\lambda_m}+ \sum\limits_{k=1}^N\left(1-\frac{\lambda_k}{\lambda_m}\right)
\oint\limits_{\partial D_k}x \frac{\partial T_k}{\partial n}\, ds=-\frac{A \cos \theta}{\lambda_m}+
\sum\limits_{k=1}^N\left(1-\frac{\lambda_k}{\lambda_m}\right) \iint\limits_{D_k} \frac{\partial T_k}{\partial x}\, dxdy,
$$
$$
T_2=-\frac{A \sin \theta}{\lambda_m}+ \sum\limits_{k=1}^N \left(1-\frac{\lambda_k}{\lambda_m}\right)\iint\limits_{D_k} \frac{\partial T_k}{\partial y}\, dxdy.
$$
Combining these values together with use of (\ref{flux_funs}) and the mean value theorem for harmonic functions, we have
$$
T_1- \imath T_2=\frac{-A e^{-\imath\theta}}{\lambda_m}+2 \sum\limits_{k=1}^N\frac{\lambda_m-\lambda_k}{\lambda_m+\lambda_k}
\iint\limits_{D_k}\,\psi_k(z)  \,dxdy=\frac{-A e^{-\imath\theta}}{\lambda_m}-2 \pi \sum\limits_{k=1}^N \rho_k  r_k^2
\,\psi_k(a_k).
$$

\vspace{3mm}

\end{document}